\documentclass[preprint,showpacs,preprintnumbers,amsmath,amssymb]{revtex4}

\usepackage{graphicx}
\usepackage{dcolumn}
\usepackage{bm}

\begin{document}

\title{$C$ axis lattice dynamics in Bi-based cuprate superconductors}
\date{\today}
\author{N.N. Kovaleva}
\altaffiliation[Also at]{\ the Institute for Solid State Physics, Russian Academy of Sciences,
Chernogolovka, Moscow distr., 142432 Russia}
\author{A.V. Boris}
\altaffiliation[Also at]{\ the Institute for Solid State Physics, Russian Academy of Sciences,
Chernogolovka, Moscow distr., 142432 Russia}
\author{T. Holden}
\author{C. Ulrich}
\affiliation{Max-Planck-Institut f\"{u}r Festk\"{o}rperforschung, Heisenbergstrasse 1,
D-70569 Stuttgart, Germany}
\author{J.L. Tallon}
\affiliation{McDiarmid Institute and Industrial Research Laboratory, Lower Hutt
New-Zealand}
\author{D. Munzar}
\affiliation{Institute of Condensed Matter Physics, Faculty of Science, Masaryk University,
Kotl\'{a}\v{r}sk\'{a} 2, CZ-61137 Brno, Czech Republic}
\author{A.M. Stoneham}
\affiliation{University College London, London WC1E 6BT, United Kingdom}
\author{B. Liang}
\author{C.T. Lin}
\author{B. Keimer}
\author{C. Bernhard}
\affiliation{Max-Planck-Institut f\"{u}r Festk\"{o}rperforschung, Heisenbergstrasse 1,
D-70569 Stuttgart, Germany}

\begin{abstract}
We present results of a systematic study of the $c$ axis lattice dynamics in
single layer Bi$_{2}$Sr$_{2}$CuO$_{6}$ (Bi2201), bilayer Bi$_{2}$Sr$_{2}$CaCu%
$_{2}$O$_{8}$ (Bi2212) and trilayer Bi$_{2}$Sr$_{2}$Ca$_{2}$Cu$_{3}$O$_{10}$
(Bi2223) cuprate superconductors. Our study is based on both experimental
data obtained by spectral ellipsometry on single crystals and theoretical
calculations. The calculations are carried out within the framework of a
classical shell model, which includes long-range Coulomb interactions and
short-range interactions of the Buckingham form in a system of polarizable
ions. Using the same set of the shell model parameters  for Bi2201, Bi2212
and Bi2223, we calculate the frequencies of the Brillouin-zone center phonon
modes of A$_{2u}$ symmetry and suggest the phonon mode eigenvector patterns.
We achieve good agreement between the calculated A$_{2u}$ eigenfrequencies
and the experimental values of the $c$ axis TO phonon frequencies which
allows us to make a reliable phonon mode assignment for all three Bi-based cuprate
superconductors. 
We also present the results of our shell model calculations for 
the $\Gamma $-point A$_{1g}$ symmetry modes in Bi2201, Bi2212 and Bi2223 
and suggest an assignment that is based on the published experimental Raman spectra. 
The superconductivity-induced phonon anomalies recently observed in the $c$ axis 
infrared and resonant Raman scattering spectra in trilayer Bi2223 are consistently explained
with the suggested assignment.  
\end{abstract}
\date{\today}
\maketitle

\section{introduction}

The unusual $c$ axis charge transport in the cuprate high-$T_{c}$
superconductors is still a subject of a controversial discussion. Of
particular interest is a sizable absorption peak that develops below T$_c$ 
in the far-infrared (FIR) range, accompanied with the strongly anomalous temperature dependence 
of some of the $c$-polarized IR phonon modes, in compounds that contain 
more than one CuO$_{2}$ plane per unit cell. Recently a reasonable 
description of these effects has been obtained with the so-called 
Josephson-superlattice model \cite{Munzar1,Gruninger,Marel,Bernhard,Zelezny1,Munzar2}. 
Here it is assumed that the individual CuO$_{2}$ planes are only weakly coupled by Josephson 
currents in the superconducting state. For bilayer compounds like 
YBa$_{2}$Cu$_{3}$O$_{7-\delta }$ (Y123) and Bi$_{2}$Sr$_{2}$CaCu$_{2}$O$_{8}$ 
(Bi2212) this results in two kinds of Josephson junctions with different
longitudinal plasma frequencies, and their out-of-phase oscillation gives
rise to a transverse Josephson plasma resonance, which has been assigned
to the absorption peak that develops below T$_c$; the Josephson currents lead 
to a modification of the dynamical local electric fields and thus are at the heart 
of the observed phonon anomalies \cite{Munzar1,Bernhard,Zelezny1,Munzar2}. 
The trilayer compounds provide an interesting testing ground for the various models 
that have been put forward to explain the unusual $c$ axis charge dynamics and the anomalous
temperature dependence of some of the IR-active $c$-polarized phonon modes
such as observed in the bilayer compounds.
Indeed, very recent experimental study of the $c$ axis optical conductivity of 
a trilayer Bi$_{2}$Sr$_{2}$Ca$_{2}$Cu$_{3}$O$_{10}$ (Bi2223) provides clear evidence 
that the transverse Josephson plasma resonance is a universal feature of the multilayer
high-T$_c$ cuprate compounds \cite{Boris}. It has been shown that the Josephson-superlattice model 
\cite{Munzar1,Bernhard, Zelezny1,Munzar2} allows qualitative description of the transverse Josephson 
plasma resonance and related phonon anomalies in Bi2223, which suggest that the Josephson currents 
lead to a strong variation of the dynamical local electric field even between the inner and the outer 
CuO$_2$ planes of a trilayer.    

The assignment of the IR-active $c$-polarized phonon modes and their 
polarization diagrams are of great importance for understanding these phenomena.
While the properties of the IR-active
phonon modes are indeed well known for the Y123 system \cite{Kress,Liu,Humlicek,Henn}, 
the situation is not that clear for the bilayer high-$T_{c}$ compounds of
the type A$_{2}$Sr$_{2}$CaCu$_{2}$O$_{8}$ (A2212; A = Tl, Bi). Despite of
experimental \cite{Zelezny1,Munzar2,Tajima,Tsvetkov} and theoretical 
\cite{Prade,Chun,Jia} efforts, the assignment of the $c$ axis IR phonon
modes in Bi2212 and Tl2212 remains contradictory, as discussed in 
\cite{Tsvetkov}. The situation is even less satisfactory for the trilayer 
high-$T_{c}$ compounds A$_{2}$Sr$_{2}$Ca$_{2}$Cu$_{3}$O$_{10}$ (A2223) where the
lack of single crystals of sufficient size and quality has inhibited
detailed experimental investigations. Shell model calculations have been
carried out for Tl2223 \cite{Kulkarni} and IR measurements have been performed
on ceramic samples of Tl2223 \cite{Zetterer} and more recently on oriented
ceramics of Bi2223 \cite{Petit}. The situation has changed only very
recently when some of the present authors managed to grow high quality
single crystals of the trilayer Bi-compound Bi$_{2}$Sr$_{2}$Ca$_{2}$Cu$_{3}$O$_{10}$ 
that are suitable for IR measurements with the light polarization
along the $c$ axis direction \cite{Liang23}. 

This manuscript reports a systematic investigation of the $c$ axis lattice
dynamics of the bismuth-based compounds Bi$_{2}$Sr$_{2}$Ca$_{n-1}$Cu$_{n}$O$_{2n+4+d}$ 
that contain either n = 1, 2, or 3 CuO$_{2}$ layers per unit cell. 
Our approach combines theoretical shell model based 
calculations with ellipsometric measurements of the $c$ axis dielectric
response on well characterized single crystals of high quality \cite{Liang23,Liang12,Liang01}. 
From the ellipsometry experiments we obtain the 
frequencies of the $c$ axis polarized IR-active phonon modes as well as
the dielectric constants. 
On the theoretical side we model the Bi-based compounds with n = 1, 2, and 3 
in the {\it I4/mmm} space group using the shell model approach 
suggesting that screening due to free charge carriers should not have major effects \cite{MStoneham}. 
This approach has been successfully applied to study lattice
dynamics and various physical properties of high-$T_{c}$ superconductors 
\cite{Prade,Kulkarni,Baetzold,Islam}. The calculations have been performed
using the GULP code \cite{Gale}. The shell model parameters have been
derived by fitting to the crystal structures in equilibrium as well as to
experimental values of the dielectric constants and frequencies of the
transverse optical (TO) phonon modes. Using the same set of shell model parameter
for the n = 1, 2, and 3 compounds we calculate the eigenfrequencies of the
Brillouin-zone center phonon modes of A$_{2u}$ symmetry and derive the
phonon mode eigenvector patterns. 
Our results for the bilayer compound Bi2212 are consistent with
those of the earlier shell model calculations in the {\it I4/mmm} 
space group by Prade {\it et al.} \cite{Prade}. 
We make an assignment of the $c$ axis polarized phonon modes observed 
in the ellipsometric spectra of single, bi-, and trilayer Bi-based 
compounds by comparing with the calculated eigenfrequencies 
of A$_{2u}$ symmetry modes. In addition, we also
show the displacement patterns and calculated eigenfrequencies for the
Raman-active phonon modes of A$_{1g}$ symmetry according to the results of
the present shell model calculations.

\section{CRYSTAL STRUCTURE AND $\Gamma $-POINT PHONONS IN 
$\mathrm{Bi_2Sr_2Ca_{n-1}Cu_nO_{2n+4+\delta}}$ WITH n = 1, 2 and 3}

As a first approximation, a body-centered-tetragonal
structure ($I4/mmm$) has been most frequently used in the interpretation of
the IR and Raman spectra of Tl- and Bi-based compounds. According to
the x-ray data \cite{Tarascon,Tarascon1}, the three Bi-based cuprates of 
general formula Bi$_{2}$Sr$_{2}$Ca$_{n-1}$Cu$_{n}$O$_{2n+4+\delta}$ 
with n = 1, 2 and 3 have similar structures of the tetragonal $I4/mmm$ space group 
($a_{tetr}$ = 3.814 \AA ), with two formula units (two primitive unit cells) 
in the crystallographic cell, as illustrated in Figure 1. 
Primitive unit cells of single, bi- and trilayer Bi-based cuprates 
differ only in the number (n-1) of CuO$_2$-Ca-CuO$_2$ slabs packed along the $c$ axis; 
upon insertion of one and two slabs the $c$ axis parameter of the crystallographic cells 
increases from 24.6 to 30.6, and 37.1 \AA \ \cite{Tarascon1}. 
The Wyckoff positions of the atoms and their site symmetries in tetragonal 
$I4/mmm$ (D$_{4h}^{17}$) primitive unit cells of 
Bi$_2$Sr$_2$CuO$_6$ (Bi2201), Bi$_2$Sr$_2$CaCu$_2$O$_8$ (Bi2212), 
and Bi$_2$Sr$_2$Ca$_2$Cu$_3$O$_{10}$ (Bi2223) are listed in Table 1, 
with the notations of the atoms given in Figure 2. 

The irreducible representations corresponding to various atomic sites in the 
three Bi-based compounds that follow from the character tables of the point 
groups are presented in the right column of Table 1. For the IR-active modes at the 
$\Gamma$-point of the Brillouin zone the displacement symmetries are A$_{2u}$
for a displacement in the $z$ direction and E$_{u}$ for displacements in the 
$x$ or $y$ directions, whereas those for the Raman-active $\Gamma$ modes
are A$_{1g}$, B$_{1g}$ and E$_{g}$, respectively. Due to full point-group
symmetry of the Cu1 (Bi2201), Cu2 (Bi2223) and Ca (Bi2212) atoms with the
(1a) site position, no Raman-active modes of these atoms are allowed, and the three
displacement vectors yield three modes of A$_{2u}$ and E$_{u}
$ (doubly degenerate) character. The oxygen atoms in the mirror cuprate
planes, O1 in single layer and O4 in trilayer Bi-compounds, of D$_{2h}$ site
symmetry, yield eigenmodes A$_{2u}$ + B$_{2u}$ + 2E$_{u}$. The in-plane
oxygen O1 with C$_{2v}$ site symmetry in the bilayer and trilayer compounds
generates eigenmodes A$_{2u}$ + A$_{1g}$ + B$_{1g}$ + B$_{2u}$ + 2E$_{g}$ +
2E$_{u}$, among them the B$_{2u}$ mode is silent and E$_{g}$ is doubly degenerated. 
The total numbers of modes grouped according to their optical activity in the three 
Bi-based compounds are summarized in Table 2. 

\begin{table}[tbp]
\caption{The irreducible representations for the atoms in tetragonal {\it I4/mmm} (D$^{17}_{4h}$)
Bi$_2$Sr$_2$CuO$_6$ (Bi2201), Bi$_2$Sr$_2$CaCu$_2$O$_8$ (Bi2212) and
Bi$_2$Sr$_2$Ca$_2$Cu$_3$O$_{10}$ (Bi2223).} 
\begin{ruledtabular}
\begin{tabular}{lllllll}    
     Comp. &Atom & Wyc.  & Site     & Irreducible & representation \\    
           &     & not.   & sym.     &             &                \\         
\colrule
Bi2201    & Cu1                & (1a) & D$_{4h}$     & A$_{2u}$              & + E$_u$                      \\
           & Bi, Sr, O2, O3     & (2e) & C$_{4v}$     & 4(A$_{2u}$ + A$_{1g}$ & + E$_g$ + E$_u$)             \\ 
          & O1                 & (2c) & D$_{2h}$     & A$_{2u}$              & + B$_{2u}$  + 2E$_u$         \\
\colrule      
Bi2212    & Ca                 & (1a) & D$_{4h}$     & A$_{2u}$              & + E$_u$                      \\
          & Bi, Sr, Cu1, O2, O3& (2e) & C$_{4v}$     & 5(A$_{2u}$ + A$_{1g}$ & + E$_g$  + E$_u$)            \\ 
          & O1                 & (4g) & ${C^v}_{2v}$ & A$_{2u}$ + A$_{1g}$ + B$_{1g}$ & + B$_{2u}$ + 2E$_g$  + 2E$_u$\\
\colrule
Bi2223    & Cu2                     & (1a) & D$_{4h}$     & A$_{2u}$ &  + E$_u$                              \\
          & O4                      & (2c) & D$_{2h}$     & A$_{2u}$ & + B$_{2u}$  + 2E$_u$                  \\
          & Bi, Sr, Ca, Cu1, O2, O3 & (2e) & C$_{4v}$     & 6(A$_{2u}$ + A$_{1g}$ & + E$_g$  + E$_u$)        \\ 
          & O1                      & (4g) & ${C^v}_{2v}$ & A$_{2u}$ + A$_{1g}$ + B$_{1g}$ & + B$_{2u}$ + 2E$_g$ + 2E$_u$ \\
\end{tabular}
\end{ruledtabular} 
\end{table}

\begin{table}[tbp]
\caption{Modes classification from the irreducible representations for
Bi2201, Bi2212 and Bi2223 in tetragonal {\it I4/mmm} (D$^{17}_{4h}$)
space group.}
\begin{ruledtabular}
\begin{tabular}{lllll}    
     Comp. & $\Gamma_{IR}$ & $\Gamma_{Raman}$ & $\Gamma_{acoustic} $ & $\Gamma_{silent}$ \\    
\colrule
Bi2201    & 5A$_{2u}$ + 6E$_u$ & 4A$_{1g}$ + 4E$_g$ & A$_{2u}$ + E$_u$ & B$_{2u}$ \\
Bi2212    & 6A$_{2u}$ + 7E$_u$ & 6A$_{1g}$ + 7E$_g$ + B$_{1g}$ & A$_{2u}$ + E$_u$ & B$_{2u}$ \\
Bi2223    & 8A$_{2u}$ + 10E$_u$ & 7A$_{1g}$ + 8E$_g$ + B$_{1g}$ & A$_{2u}$ + E$_u$ &  2B$_{2u}$  
\end{tabular}
\end{ruledtabular} 
\end{table}

\begin{table}[tbp]
\caption{The irreducible representations for the atoms in orthorhombic {\it Amaa} 
(D$^{20}_{2h}$) Bi$_4$Sr$_4$Cu$_2$O$_{12}$ (Bi2201)and Bi$_4$Sr$_4$Ca$_2$Cu$_4$O$_{16}$ (Bi2212).}
\begin{ruledtabular}
\begin{tabular}{lllll}    
    Comp. &Atom                & Wyc.      & Irreducible  representations \\    
          &                    & not.       &                             \\         
\colrule
Bi2201    & Cu1                & (2e) & A$_u$ + 2B$_{1u}$ + 2B$_{2u}$ + B$_{3u}$   \\
          &                    &      &B$_{1u}$ + B$_{2u}$ tetragonal silent\\   
\colrule          
          & Bi, Sr, O2, O3     & (4l) & 4(A$_u$ + 2B$_{1u}$ + 2B$_{2u}$ + B$_{3u}$ + 2A$_g$ + B$_{1g}$ + B$_{2g}$ + 2B$_{3g}$) \\ 
          &                    &      & 4(B$_{1u}$ + B$_{2u}$ + A$_g$ + B$_{1g}$ + B$_{3g}$) tetragonal silent\\  
\colrule
          & O1                 & (4h) & A$_u$ + B$_{1u}$ + 2B$_{2u}$ + 2B$_{3u}$ + A$_g$ + B$_{1g}$ + 2B$_{2g}$ + 2B$_{3g}$ \\
          &                    &      &A$_g$ + B$_{1g}$ + 2B$_{2g}$ + 2B$_{3g}$ tetragonal silent\\
\colrule      
Bi2212    & Ca                 & (2e) & A$_u$ + 2B$_{1u}$ + 2B$_{2u}$ + B$_{3u}$ \\
          &                    &      &B$_{1u}$ + B$_{2u}$ tetragonal silent\\ 
\colrule
          & Bi, Sr, Cu1, O2, O3& (4l) & 5(A$_u$ + 2B$_{1u}$ + 2B$_{2u}$ + B$_{3u}$ + 2A$_g$ + B$_{1g}$ + B$_{2g}$ + 2B$_{3g}$)\\             
          &                    &      & 5(B$_{1u}$ + B$_{2u}$ + A$_g$ + B$_{1g}$ + B$_{3g}$) tetragonal silent\\  
\colrule
          & O11,O12            & (4h) & 2(A$_u$ + B$_{1u}$ + 2B$_{2u}$ + 2B$_{3u}$ + A$_g$ + B$_{1g}$ + 2B$_{2g}$ + 2B$_{3g})$\\
          &                    &      & B$_{1u}$ + 2B$_{2u}$ + 2B$_{3u}$ + A$_g$ + B$_{1g}$ + 2B$_{2g}$ + 2B$_{3g}$ tetragonal silent\\  
\end{tabular}
\end{ruledtabular} 
 \end{table}

It has been shown that the actual structure of Bi-based compounds is better
represented by orthorhombic {\it Cccm} (D$^{20}_{2h}$) \cite{Imai,Torardi,Sequeira,Ito} 
or {\it Ccc2} (C$^{13}_{2v}$)) \cite{Bordet,Bellingeri} space groups, 
most of the structural studies verify that the orthorhombic $Cccm$ structure is correct. 
To follow the consequences of the lowering symmetry it is convenient to use 
the {\it Amaa} space group, which is a nonconventional setting of $Cccm$.
The orthorhombic {\it Amaa} structure has two formula units per primitive 
face centered cell, with in-plane lattice parameters $a$ and $b$ of 
approximately $a_{tetr}\sqrt{2}$, and four formula units 
for crystallographic cell, with the same $c$ axis parameter. 
By way of example, in the right column of Table 3 we present 
the irreducible representations corresponding to various atomic sites in 
the orthorhombic $Amaa$ Bi$_4$Sr$_4$Cu$_2$O$_{12}$ (Bi2201) and 
Bi$_4$Sr$_4$Ca$_2$Cu$_4$O$_{16}$ (Bi2212) according to the structural data of 
Imai {\it et al.} \cite{Imai}, where we also show the modes which are silent 
in the tetragonal symmetry. 
The total numbers of modes grouped according to their optical activity in the
$Amaa$ Bi2201 and Bi2212 are summarized in Table 4. 
In Table 5 we show components of the IR and Raman-active normal modes of
the atoms with different Wyckoff positions at the $\Gamma$-point 
of the Brillouin zone in the orthorhombic {\it Amaa} space group.
From Tables 1 and 3 one can see that if the {\it I4/mmm} symmetry breaks down 
owing to the orthorhombic distortion of the structure, the degeneracy is removed. 
Thus, for each atom with (2e) site position in the {\it I4/mmm} symmetry 
(Bi, Sr, O2, O3 and Cu1 in Bi2212) normal modes A$_{2u}$, A$_{1g}$, 
one E$_u$, and one E$_g$ will become nondegenerate and, as a result, split into two; 
in addition, one new mode of B$_{1g}$ character will appear 
in the orthorhombic {\it Amaa} structure.  
For the atoms Cu1 in Bi2201 and Ca in Bi2212 two new IR-active modes 
of B$_{1u}$ and B$_{2u}$ character are expected in the {\it Amaa} structure. 
Due to breaking of the tetragonal symmetry the O1 atom in Bi2201 will become 
"Raman-active", and, as a result, a number of Raman-active modes of A$_g$, B$_{1g}$, 
2B$_{2g}$, and 2B$_{3g}$ character will appear.  
\begin{table}[tbp]
\caption{Modes classification from the irreducible representations for
Bi2201 and Bi2212 in orthorhombic {\it Amaa} space group.}
\begin{ruledtabular}
\begin{tabular}{lllll}    
     Comp. & $\Gamma_{IR}$ & $\Gamma_{Raman}$ & $\Gamma_{acoustic} $ & $\Gamma_{silent}$ \\    
\colrule
Bi2201   & 10B$_{1u}$ + 11B$_{2u}$ +6B$_{3u}$ & 9A$_g$ + 5B$_{1g}$ + 6B$_{2g}$ + 10B$_{3g}$ & B$_{1u}$ + B$_{2u}$ + B$_{3u}$ & 6A$_u$ \\
Bi2212   & 13B$_{1u}$ + 15B$_{2u}$ +9B$_{3u}$ & 12A$_g$ + 7B$_{1g}$ + 9B$_{2g}$ + 14B$_{3g}$& B$_{1u}$ + B$_{2u}$ + B$_{3u}$ & 8A$_u$ \\
\end{tabular}
\end{ruledtabular} 
\end{table}

\begin{table}[tbp]
\caption{Components of normal modes of atoms with different Wyckoff positions
in orthorhombic {\it Amaa} space group.}
\begin{ruledtabular}
\begin{tabular}{lllllllll}
Wyc. not. & A$_g$ & A$_u$ & B$_{3g}$ & B$_{3u}$ & B$_{1g}$ & B$_{1u}$ &  B$_{2g}$ & B$_{2u}$ \\   
\colrule
(2e)       &     & $x$&     & $x$&    & $zy$&     & $yz$\\
(4l)       & $zy$& $x$& $yz$& $x$& $x$& $zy$& $x$ & $yz$\\
(4h)       & $z$ & $z$& $xy$&$xy$&$z$ & $z$ &$yx$ & $yx$
\end{tabular}
\end{ruledtabular} 
\end{table}

However, it has been known that the actual structure of Bi-based cuprates is only
pseudo-orthorhombic due to stacking faults, extra-stoichiometric oxygen and
incommensurate superstructural modulations along the $b$ axis 
in BiO and SrO layers \cite{Jacubowicz}, 
which result in substantial oxygen and cation disorder. These factors can
yield additional phonon modes, in particular, "disorder-induced" modes.
The crystal structure of Bi-based compounds therefore should be described
in terms of a larger elementary cell and lower crystal symmetry. 
The origin of the superstructural incommensurate modulations of the atoms in
BiO and SrO layers is not fully understood. The most common assumption is
that the modulations are associated with extra-stoichiometric
oxygen in the bilayer Bi$_{2}$O$_{2+\delta}$ that affects the orthorhombicity 
in the Bi-based cuprates. 
In this situation, whatever approximation either tetragonal or orthorhombic is 
adopted to make an assignment in the experimental phonon spectra, 
the first problem one has to solve is to distinguish the extra phonon modes 
- beyond the considered symmetry - due to effects of superstructural incommensurate 
modulations and presence of extra-stoichiometric oxygen in the Bi-based compounds. 

Since the deviation from tetragonality in Bi-based compounds is not large we suggest to start with 
the tetragonal $I4/mmm$ approximation assuming that modes silent in tetragonal symmetry will be weak, 
and the simple tetragonal $I4/mmm$ approximation should provide a useful first approximation for the
assignment of the most intense IR and Raman-active phonon modes. To explain more details in 
the experimental spectra due to possible orthorhombicity effects we suggest to invoke 
the group theory analysis based on the orthorhombic $Amaa$ (D$^{20}_{2h}$) space group (see Table 3).  
 
\section{samples and experimental technique}

The Bi2201, Bi2212 and Bi2223 single crystals have been grown by the travelling
solvent floating zone (TSFZ) method \cite{Liang23,Liang12,Liang01}. The feed
rods with the nominal composition of Bi$_{2.1}$Sr$_{1.9}$Ca$_{n-1}$Cu$_{n}$O$_{2n+4+\delta}$ 
(n = 1, 2 and 3) were prepared by the conventional solid state
method. The slow growth rates of 0.5 and 0.2 mm/h were used for the Bi2201
and Bi2212 crystals, respectively. Since the Bi2223 crystallization field is
the narrowest among the three members of Bi-based family, an extremely slow
growth rate of 0.04 mm/h was used. The Bi2201, Bi2212 and Bi2223 crystals
were grown under applied oxygen pressure $p(O_{2}$), in air, and in mixed
gas flow of 80\% Ar and 20\% O$_{2}$, respectively. Because of the
significant difference between the lattice parameters of the $a$ and $b$ axis
as compared to the one along the $c$ axis, the growth rate is highly
anisotropic and very slow along $c$. One typically obtains platelike crystals
with large $ab$ planes of about 6 $\times $2 mm$^{2}$ but with much smaller
dimensions along the $c$ axis. The typical thickness is 1.5 $\div $ 2 mm for
Bi2201, about 1 mm for Bi2212, less than 1 mm for Bi2223.

The crystals were characterized by energy dispersive X-ray (EDX) analysis,
X-ray diffraction (XRD), transmission electron microscopy (TEM), magnetic
susceptibility and resistivity measurements. The as-grown Bi2212 crystals
are phase-pure with a pseudotetragonal structure, they are almost optimally
doped with T$_{c}$ = 91 K and $\Delta T_{c}$ = 2 K. The as-grown Bi2223
crystal contains 95-98 \% of the Bi2223 pseudotetragonal phase with a minor
fraction of layer-intercallated Bi2212. In the as-grown state the Bi2223
crystal was underdoped with T$_{c}$ = 97 K (midpoint) and $\Delta
T_{c}$ (10-90\% shielding fraction) = 7 K. After a post-annealing treatment
for ten days in flowing oxygen at 500 $^{\circ }$C with a subsequent rapid
quenching to room temperature, the crystal became nearly optimally doped,
with T$_{c}$ = 107 K and $\Delta T_{c}$ = 3 K. The undoped Bi2201 crystals
grown by the TSFZ method are often nonsuperconducting. The superconductivity
of the as-grown Bi2201 crystals is critically controlled by the oxygen
content, therefore the growth was carried out under oxygen pressures $p(O_{2}$) 
ranging from 1 to 10 bar. The T$_{c}$ of the as-grown Bi2201 crystals,
determined from magnetic susceptibility measurements varied between 3.5 and
8.5 K upon the Bi/Sr ratio and the oxygen pressure $p(O_{2}$) applied in the
course of growing. The structure of the as-grown Bi2201 crystals was
determined to be pseudotetragonal, the lattice parameters were found to
dependent upon the Bi/Sr ratio in the crystal composition. A pair of
as-grown Bi2212 single crystals (taken from the same part of an ingot) was
used for oxygen isotope replacement. The selected samples were annealed in
isotopically enriched oxygen under identical conditions following a similar 
procedure described in Ref. \cite{Tallon}. 

For the ellipsometric measurements the crystal surfaces were polished to
optical grade. The technique of ellipsometry provides significant advantages
over conventional reflection methods in that (i) it is self-normalizing and
does not require reference measurements and (ii) $\epsilon _{1}(\omega )$ and 
$\epsilon _{2}(\omega )$ are obtained directly without a Kramers-Kronig
transformation (see Refs. \cite{Bernhard,Henn1} for a description of the
technique). The ellipsometric measurements have been performed at the U4IR
beamline of the National Synchrotron Light Source (NSLS) at Brookhaven
National Laboratory, USA and at the infrared beamline of the synchrotron
radiation source ANKA at Karlsruhe Research Center, Germany. Home-built
ellipsometers attached either to a Nicolet Magna 860 or a "Bruker" IFS
66v/S FT-IR spectrometer were used. The high brilliance of the synchrotron
light source due to the small beam divergence enabled us to perform very
accurate ellipsometric measurements in the far-IR range even on samples with
comparably small $ac$ faces.

\section{$C$ AXIS IR SPECTRA OF $\mathrm{Bi}$2201, $\mathrm{Bi}$2212, AND $%
\mathrm{Bi}$2223 SINGLE\ CRYSTALS} 

Figure 3 shows the real parts of the $c$ axis optical conductivity $\sigma
_1(\omega )$ and the dielectric function $\epsilon _1(\omega )$ 
at 150 K for the Bi2212 and Bi2223 single crystals with $T_{c}$ =
91 K and $T_{c}$ = 107 K. It is evident that the optical spectra look rather
similar for both compounds. This is a consequence of the generic structural
features of these two compounds. The electronic background is extremely weak and 
the spectra are dominated by the contributions of several IR-active phonon modes. 
Using the classical dispersion analysis we
fit a set of Lorentzian oscillators simultaneously to $\epsilon _{1}(\omega )
$ and $\epsilon _{2}(\omega )$ in the investigated spectral range. To
describe the weak and rather flat electronic background in a Kramers-Kronig
consistent way we have also included a sum of a Drude term and several very
broad Lorentzian oscillators. As for the present purpose this description is
rather formal we only limited the half-widths of the broad Lorenzians to
values between 150 and 500 cm$^{-1}$. In Table 6 we list the frequencies $\nu _{j}$, 
oscillator strengths $S_{j}$, and damping parameters $\gamma _{j}$ 
of the TO phonon bands observed in the $\epsilon_2(\omega)$ spectra of Bi2212
and Bi2223 single crystals, together with determined values of the high-frequency
dielectric constants, $\epsilon_{\infty}$.

\begin{table}[tbp]
\caption{Frequencies $\nu _{j}$ (in cm$^{-1}$), oscillator strengths S$_{j}$, 
and damping parameters $\gamma _{j}$ (in cm$^{-1}$) of the TO phonon bands 
observed in the $c$-polarized $\epsilon _{2}(\omega )$ spectra (T = 150 K) 
of the single crystals Bi2201, Bi2212, and Bi2223. 
The assignment with the calculated frequencies of A$_{2u}$ $(I4/mmm)$ eigenmodes is given.}
\begin{ruledtabular}
\begin{tabular}{lllllllll} 
Bi2212 & & & Bi2223  & & & Bi2201  & & \\
$\nu_j (S_j)$ & $\gamma_j$ & A$_{2u}$ & $\nu_j (S_j)$ & $\gamma_j$ & A$_{2u}$ & $\nu_j (S_j)$ & $\gamma_j$ & A$_{2u}$  \\    
\colrule
 97       &  --  & 93   & 97           & --   & \ 88 & 109 (0.56) & 5.00 &  \ 77   \\ 
 --       &  --  & --   & 127 (0.10)   & 9.1  & 145  &  --        &  --  &  --     \\
168 (0.29)& 8.7  & 195  & 170 (0.63)   & 11.2 & 211  & 165 (0.45) & 12.3 & 175     \\ 
210 (0.34)& 12.7 & 304  & 212 (0.77)   & 21.4 & 252  & 208 (0.65) & 14.5 &  --     \\ 
282 (0.16)& 12.3 & --   & 277 (0.50)   & 17.4 &  --  & --         &  --  &  --     \\ 
304 (1.13)& 21.7 & 343  & 305 (1.43)   & 27.6 & 335  & 301 (1.31) & 24.7 & 293     \\   
358 (0.33)& 30.7 & 403  & 360 (0.46)   & 30.7 & 378  & 385 (0.85) & 25.2 & 353     \\ 
 --       &  --  & --   & 402 (0.16)   & 24.0 & 516  &  --        &  --  &  --     \\ 
462 (0.02)& 10.5 & --   & 466 (0.02)   & 10.5 & --   & --         &  --  & --      \\ 
521 (0.22)& 85.3 & --   & 512 (0.26)   & 81.0 & --   & 501 (0.10) & 48.0 & --      \\ 
583 (0.65)& 38.2 & 643  & 582 (0.61)   & 41.9 & 666  & 586 (0.42) & 31.2 & 596     \\ 
640 (0.03)& 17.8 & --   & 640 (0.03)   & 17.8 & --   & 639 (0.01) & 12.5 & --      \\ 
\colrule
$\epsilon_\infty$ = 4.91 &  & & $\epsilon_\infty$ = 5.77& & &  $\epsilon_\infty$ = 4.69 & &
\end{tabular}
\end{ruledtabular} 
\end{table}
For the Bi2223 we assign the main phonon bands at 97, 127, 170, 212,
305, 360, 402, and 582 cm$^{-1}$ to the eight A$_{2u}$ character modes.
These modes are clearly identified especially due to the resonance features
in $\epsilon _{1}(\omega )$ as shown in Figure 3(b). Except for the two new
intrinsic phonon modes in Bi2223 which appear at 127 and 402 cm$^{-1}$, the 
modes are observed at similar frequencies in both compounds. 
We believe that the remaining phonon features in Figure 3 around 
277, 466, and 640 cm$^{-1}$ are beyond the tetragonal $I4/mmm$ approximation, 
as discussed in Section 2. 
The feature at 640 cm$^{-1}$ appears to be a weak satellite of the pronounced
high-frequency mode at 583 cm$^{-1}$. It is likely related to the effects of 
characteristic incommensurate modulations which induce strong distortions in 
the SrO and BiO planes. Indeed, this feature almost disappears with Pb doping 
which significantly reduces these superstructural modulations \cite{Tsvetkov,unpublished1}. 
Concerning the mode at 277 cm$^{-1}$, on the low-frequency side of the pronounced band 
at 305 cm$^{-1}$, we find that its oscillator strength correlates with the amount 
of extra-stoichiometric oxygen that is incorporated within the Bi$_{2}$O$_{2+\delta}$ 
bilayer during the oxygen annealing treatment \cite{Boris}. 
By contrast to the 640 cm$^{-1}$ satellite this feature remains persistent 
with Pb doping \cite{unpublished1}. 
As a very weak feature at around 466 $cm^{-1}$ is quite distant from 
the main phonon features we associate with the A$_{2u}$ character modes, 
we would suggest that it is due to a contribution from some in-plane 
oxygen-related (O2 or O3) B$_{2u}$ mode which has a $z(c)$ component 
(see Tables 3 and 5), or it can be a "disorder-induced" defect mode.
    
The 150 K spectra of the real parts of the $c$ axis optical
conductivity $\sigma _1(\omega)$ and dielectric function $\epsilon_1(\omega)$ 
of the as-grown nonsuperconducting Bi2201 single crystal, with the 
stoichiometric composition Bi$_2$Sr$_2$CuO$_6$, are shown in Figure 4. 
The values of the fitted frequencies $\nu _j$, oscillator strengths $S_j$, 
and damping parameters $\gamma _j$ of the TO phonon bands in the $\epsilon_2(\omega)$ 
are also listed in Table 6. To the best of our knowledge this is the first report 
of the $c$-polarized phonon spectra of the Bi2201 single crystal. 
Also we would like to note that the $c$ axis spectra of the Bi2201 reported here 
exhibit remarkable difference with respect to those of the isostructural single layer 
Tl compound Tl2201 \cite{Tsvetkov}. Six strong TO phonon features 
at around 109, 165, 208, 301, 385, and 586 cm$^{-1}$ 
are clearly identified in the $\sigma_1(\omega)$ spectrum of Bi2201, 
as shown in Figure 4(a), whereas only five A$_{2u}$ character modes are 
expected from the consideration in the ideal $I4/mmm$ space group (see Tables 1 and 2).
In addition to a weak satellite feature at around 640 cm$^{-1}$, which we have
already discussed in the previous paragraph, one "extra" mode appears 
at low frequencies in the Bi2201. 
This presumably reflects that the deviation from the tetragonality is most 
pronounced for the single layer Bi2201. We expect significant amplitudes
in the $c$-polarized IR spectrum for the tetragonal forbidden B$_{1u}$ mode
of the Bi atoms, which are known to deviate most strongly from the tetragonal symmetry. 
We can also suggest that one new B$_{1u}$ mode associated with the (2e) site position 
of the Cu1 atoms in the $Amaa$ Bi2201, which is tetragonal silent, can appear in 
the orthorhombic symmetry (see Table 3). 
Also we have found that the spectra of superconducting Bi2201 crystals 
with T$_c$ varying between 3.5 and 8.5 K depending on the Bi/Sr ratio and 
the oxygen stoichiometry exhibit almost the same phonon features.

We have furthermore investigated the oxygen isotope effect for $^{16}$O and 
$^{18}$O substitution in Bi2212 as shown in Figure 5. For a mode with
predominantly oxygen character the expected red shift of the eigenfrequency
should be proportional to [m($^{16}$O)/m($^{18}$O)]$^{1/2}$ - 1, i. e. about 5.7\%. 
A lower value of this so-called isotopic frequency shift signals that
the corresponding phonon mode contains a significant contribution from the
heavier (metal) ions. As shown in Figure 5(a), a noticeable
red shift upon $^{18}$O substitution is observed for the four A$_{2u}$ 
phonons at 210, 304, 358, and 583 cm$^{-1}$; the estimated values of the shifts
are 4.1, 4.4, 5.3, and 5.5\%, respectively. This result confirms the
expected trend that the three high frequency modes at 304, 358, and 583 cm$^{-1}$ 
have predominantly oxygen character. It is an interesting finding
that the low-frequency mode at 210 cm$^{-1}$ also exhibits a sizeable
red shift, which suggests that it contains a significant contribution from
the oxygen vibrations. 

Some of the $c$-polarized IR phonon modes in the bilayer Bi2212 and trilayer Bi2223 
compounds exhibit strongly anomalous temperature dependence at T$_c$, 
accompanied with the development of a sizable absorption peak below T$_c$ 
in the FIR range \cite{Munzar1,Zelezny1,Munzar2,Boris} . 
Figures 6(a,b) show the real part $\sigma_{1}(\omega)$ of the $c$ axis optical 
conductivity of the Bi2212 and Bi2223 compounds at two different 
temperatures, above $T_c$ and well below $T_c$. In the normal state the spectra 
exhibit hardly any noticeable changes, except for a sharpening of the phonons 
with decreasing temperature. Right below T$_c$, however, the spectra 
change appreciably. This is also illustrated in Figure 6(c) which displays 
the difference $\sigma_1(T << T_c,\omega) - \sigma_1(T\gtrsim T_c,\omega)$. 
The most prominent feature is the broad absorption band around 500-550 cm$^{-1}$
which appears below $T_{c}$ and grows rapidly with decreasing temperature.
This absorption band has been identified in both Bi2212 and Bi2223 and
has been attributed to the transverse Josephson-plasma resonance (t-JPR)
that is a universal feature of the multilayer high-T$_c$ cuprate compounds 
\cite{Bernhard,Munzar1,Munzar2}.
It is evident from Figure 6 that the  formation of the t-JPR is associated with 
an anomalous temperature dependence of the phonon modes at 360, 402 
(specific for trilayer Bi2223) and 582 cm$^{-1}$ denoted by $A$, $B$ and $C$ in Figure 6. 
The principal difference between the phonon anomalies observed 
in the bilayer and trilayer Bi-systems at T$_c$ is that in the latter system 
the changes of the spectral weight of the apical phonon mode are much more 
pronounced and the additional phonon mode exhibiting the opposite 
temperature dependence is observed at 402 cm$^{-1}$. 

\section{Theoretical calculations}

\subsection{Shell model approximation and potential parameters}

To estimate the frequencies at which the various eigenmodes appear and,
consequently, to make the assignment in the experimental phonon spectra (at T = 150 K), 
we have performed lattice-dynamical calculations using the shell model approximation. 
We performed the calculations for the three Bi-based structures Bi2201, 
Bi2212, and Bi2223 within the ideal tetragonal $I4/mmm$ (D$^{17}_{4h}$) space group, 
using the lattice constants and internal $z$ structural parameters 
reported in Refs. \cite{Tarascon,Tarascon1}. Screening has not been taken into account 
in this modeling. But screening should not noticeably affect the IR phonons along 
the poorly conducting $c$ axis direction which we investigate in this work.  
Moreover, according to the study of the phonon screening of the in-plane IR-active modes 
in high-T$_c$ superconductors the majority of the lattice modes have oscillator 
strength similar to those found in the undoped insulating materials \cite{Homes}. 
This surprising result is explained by the inhomogeneous charge distribution 
in the CuO$_2$ planes due to which the ionic cores are no longer effectively screened, and 
the carriers will not be able to effectively screen the induced dipole moment \cite{Homes}. 
Therefore, we apply the shell model approach suggesting that that screening 
due to free charge carriers should not have major effects \cite{MStoneham}. 
Earlier the shell model approach has been successfully applied to study lattice
dynamics and various physical properties of high-$T_{c}$ superconductors 
\cite{Prade,Kulkarni,Baetzold,Islam}. 
Our calculations are performed using the GULP code \cite{Gale}. 
In the context of the shell model, the lattice is considered as an assembly of polarizable ions, 
represented by massive point cores and massless shells coupled by isotropic harmonic 
forces. The interaction includes contributions from Coulomb and 
short-range interactions. Short-range interactions between relatively small cations are usually ignored.    
The core and shell charges and the spring constant of each ion are parameters of the model. 
The short-range potentials used for the shell-shell interactions are of the Buckingham form 
\begin{eqnarray}
 V_{ij} = A_{ij} exp {(-r/\rho_{ij})} - C_{ij}/r^6.  
\label{eq:one}
\end{eqnarray}
The Buckingham potential parameters as well as the atoms' core and shell charges and 
the relevant spring constants were fitted in this work. 
To obtain a self-consistent set of 
the shell model parameters acceptable for a family of Bi-based high-T$_c$ cuprates, 
we performed fits simultaneously to the structures of Bi2201, Bi2212, and Bi2223. 
As the starting values for the shell model parameters -- pair potentials, net ionic charges, 
shell and core charges and relevant spring constants -- we used the ones obtained 
by Baetzold \cite{Baetzold} for a bilayer cuprate YBa$_2$Cu$_3$O$_{7-\delta}$ system. 
We have two reasons to keep the net ionic charges of the atoms in the cuprate planes 
of +2$\mid e \mid$ for Cu1 and Cu2, and -2$\mid e \mid$ for O1 and O4 . 
First, it is in line with the approach of Baetzold for Y123 \cite{Baetzold} 
and second, this is required if we want to describe single, bi- and 
trilayer Bi-systems by the same set of the potential parameters. 
Otherwise, by dividing the non-integer charges between
the cuprate planes we would have different pair potential parameters in each
case. Following the approach of Baetzold, we also keep the same effective
charges of -1.67$\mid e\mid $ for O2 and O3. We take the formal ionic charge
+2$\mid e\mid $ for Sr, and to satisfy the electroneutrality conditions we
have to assume an effective charge +2.33$\mid e\mid $ for Bi. 
The formal charges here describe the dipole moment per unit displacement, 
rather than the charge within some region of space. 
It was noted by Catlow and Stoneham that even the integer 
charge-distribution method can be successfully employed to accurately calculate 
physical properties of solids where ionicity may vary over a range of values \cite{Stoneham}.
\begin{table} 
\caption
{Potential parameters for short-range shell-shell interactions (Eq. (1)),
ion shell charges ($Y$) and force constants ($k$) in tetragonal {\it I4/mmm} (D$^{17}_{4h}$) 
Bi$_2$Sr$_2$CuO$_6$, Bi$_2$Sr$_2$CaCu$_2$O$_8$, 
and Bi$_2$Sr$_2$Ca$_2$Cu$_3$O$_{10}$; 
in-plane O1 and O4 oxygen ions are taken of a net charge -- 2.0 $\mid e \mid$, 
out-of-plane O2 and O3 oxygen ions are taken of a net charge -- 1.67 $\mid e \mid$, 
$r_{cutoff}$ = 99\AA.}
\begin{ruledtabular}
\begin{tabular}{llrlllr}    
   &  &  A(eV) &  &  $\rho $(\AA) &  & C(eV$\cdot $\AA $^{-6}$) \\    
\colrule
Cu1-O1; Cu2-O4 & &3889.8     & &0.24273 & & 0.00 \\
Cu1-O2,O4; Cu2-O1 & &2708.0  & &0.22074 & & 0.00 \\
Ca-O1; Ca-O4  & & 14882.2    & &0.24203 & & 0.00 \\  
Sr-O1  & &1487.9   & &0.33583 & & 0.00 \\
Sr-O2  & &20097.2  & &0.24823 & & 0.00 \\
Sr-O3  & &30202.0  & &0.24823 & & 0.00 \\
Bi-O2  & & 15671.0 & &0.22074 & & 0.00 \\  
Bi-O3  & &147997.9 & &0.22074 & & 0.00 \\  
Cu1-Sr & &167925.4 & &0.22873 & & 0.00 \\  
Sr-Sr  & &3680.5   & &0.25880 & & 0.00 \\  
Sr-Bi  & &158000.0 & &0.22873 & & 0.00 \\
Bi-Bi  & &7100.0   & &0.22873 & & 0.00 \\
Ca-Ca  & &4500.0   & &0.22873 & & 0.00 \\
O-O  & &22764.0   & &0.14900  & &75.00 \\
\\ 
  & Ion &  & Y ($\mid e \mid$) &  &k (eV $\cdot $\AA  $^{-2}$) \\    
\colrule
  & Ca  &  &\ 2.00     & &99.0 \\
  & Cu1, Cu2   &  &\ 2.00     & &999999.0 \\
  & Bi  &  &\ 2.33334  & &27.2 \\
  & Sr  &  &\ 5.09995  & & 146.7 \\
  & O1  & &-3.25760    & &26.0 \\
  & O4  & &-3.25760    & &76.0 \\
  & O2, O3 & &-3.25760 & &100.0 \\  
\end{tabular}
\end{ruledtabular} 
\end{table} 
We assume the shell charge of -3.2576$\mid e \mid$ for all oxygen ions, as in \cite{Baetzold}. 
The Sr, Bi, Ca and O ions are 
treated as polarizable, while the Cu ions are treated as unpolarizable. 
The fitting procedure keeps O-O (shell-shell) interaction as those of 
typical oxides \cite{Catlow,Popov}. 
The structures were allowed to relax to equilibrium conditions 
under the symmetry ($I4/mmm$) restrictions, the cores and shells were allowed to move separately. 
The forces on the atoms were minimized by varying the potential parameters, 
the eigenvalues for the first three modes for all three compounds had approximately 
zero values, indicating a stable structure in each case. 
Normally a good fit requires some information on physical properties - 
elastic, dielectric, and piezoelectric (where applicable) constants - as well 
as the structure. 
In general, vibrational frequencies contain far more information than any of the above. 
However, in fitting the frequencies a reasonable implication for the assignment 
has to be suggested, at least for some key modes. 
In the present work a final fit was done to the available experimental data 
on the bilayer Bi2212 compound.
The criterion of a successful fit was good agreement with the structural parameters \cite{Tarascon}, 
provided that approximately zero forces are applied for all atoms, 
with the $c$ axis static $\epsilon_0$ and high-frequency $\epsilon_{\infty}$ 
dielectric constants, and with the frequencies of the $c$ axis IR- and Raman-active 
optical phonons \cite{remark}. 
The resulting Buckingham potential parameters, ionic shell charges ($Y$) 
and force constants ($k$) are presented in Table 7. 
To simulate different in-plane and out-of-plane bond lengths in the structures 
of Bi-based compounds, as well as anisotropy of oxygen polarizability, 
generally we have arrived at different pair potentials for the Cu-O, Sr-O, 
and Bi-O in-plane and out-of-plane short-range interactions. 
It is particularly remarkable that the pair potentials for the Bi-O2 and Bi-O3 
short-range interactions are significantly different, and this fact seems to be 
mainly due to very different Bi-O bond lengths, i.e., in the structures of Bi-based 
compounds we have the Bi-O2 bond length of 2.22 $\AA$, the in-plane Bi-O3 bond length 
of 2.71 $\AA$, and the longest Bi-O3 bond length, between weakly coupled 
Bi$_2$O$_2$ biplanes, of 2.97 $\AA$ \cite{Tarascon}. 
\begin{table} 
\caption
{A comparison between the experimental and calculated structures of Bi2212,
Bi2223, and Bi2201 ({\it I4/mmm} (D$^{17}_{4h}$)). 
Experimental values of the fractional atomic coordinates and corresponding 
calculated values are given, with the experimental errors for Bi2212 structure 
shown in brackets.}  
\begin{ruledtabular}
\begin{tabular}{lllllll}  
$\ $    &     Bi2212&          & Bi2223  &          &Bi2201   &           \\
$\ $    & $z$, exp. &$z$, calc.&$z$, exp.& $z$,calc.&$z$, exp.& $z$, calc. \\
\colrule
Ca      & 0.00      &  0.00    & 0.0443  & 0.0464   & --      &  --    \\
Sr      & 0.1097(5) &  0.1096  & 0.1346  & 0.1369   & 0.0759  & 0.0696 \\
Bi      & 0.3022(3) &  0.3043  & 0.292   & 0.297    & 0.3148  & 0.3175 \\
Cu1     & 0.4456(9) &  0.4334  & 0.4109  & 0.4010   & 0.00    & 0.00   \\
Cu2     &  --       & --       & 0.00    & 0.00     & --      & --     \\ 
O1      & 0.446(3)  &  0.445   & 0.411   & 0.408    & 0.00    & 0.00   \\
O2      & 0.375(4)  &  0.372   & 0.353   & 0.352    & 0.405   & 0.401  \\
O3      & 0.205(4)  &  0.211   & 0.213   & 0.219    & 0.194   & 0.201  \\
O4      &  --       & --       & 0.00    & 0.00     & --      & --     \\
\end{tabular} 
\end{ruledtabular}
\end{table}  
Thus elaborated, our shell model parameters allow us to describe all three 
Bi-based structures by the same set of short-range pair-potential parameters, 
listed in Table 7. To compare the experimental and calculated structures 
we present experimental values of the fractional atomic coordinates obtained 
from Refs. \cite{Tarascon,Tarascon1} and the corresponding calculated values 
in Table 8, where we also show the experimental errors for the Bi2212 fractional
coordinates  \cite{Tarascon} in brackets. 
As far as our final fit is based on the Bi2212 crystal, the best agreement 
with the internal structural parameters is naturally obtained for this system. 
As can be seen from Table 8 these parameters reproduce also reasonably well 
the Bi2223 and Bi2201 structures. Slightly varying the shell model parameters 
we can arrive at a better agreement with the particular structure chosen, 
although we would rather aim to obtain a self-consistent set of shell model parameters, 
which can account for the common generic features in all three Bi-based structures. 
Going ahead, we would like to mention here that this would not qualitatively 
affect the character of the eigenmodes resulting from our shell model calculations, 
although the eigenfrequencies and amplitudes of the eigenvectors of the 
$c$ axis phonon modes can slightly change. 
The elastic properties calculated with these parameters do not exhibit unphysical 
(i.e., negative values), and the eigenvalues for the first three modes have 
approximately zero values in all three compounds which is indicative of a stable structure. 
We have also arrived at a good agreement with the experimental values for dielectric 
constants. All three compounds Bi2212, Bi2223, and Bi2201 have close values 
of $c$ axis dielectric constants within the range 
$\epsilon _{0}\simeq 9.5 \ \div $ 12.0 and 
$\epsilon_{\infty } \simeq 4.9 \ \div $ 5.8, as estimated from our experimental 
$\epsilon_1(\omega)$ and $\epsilon_2(\omega)$ spectra. 
The corresponding calculated values fall in the range $\epsilon _{0}\simeq$ 
10.9 $\div$ 13.9 and $\epsilon_{\infty } \simeq 4.8 \ \div$ 5.6. 
Certainly, the obtained set of shell mode parameters is not unique. 
Nevertheless, given the fact that by using the same set of parameters 
we can reproduce the structures and main phonon bands as observed for 
all three compounds we obtain more confidence in the assignment for the phonon modes. 

\subsection{Calculated $\Gamma$-point A$_{2u}$ symmetry modes in Bi2201,
Bi2212 and Bi2223}

In Figure 7 we show the eigenvector patterns corresponding to six eigenmodes
of A$_{2u}$ symmetry and one silent mode of B$_{2u}$ symmetry for the Bi2212, 
resulting from our lattice-dynamical calculations. 
Under each patch we list values of the eigenfrequencies and below 
we make an assignment based on a comparison with the experimental frequencies 
of the TO phonon bands observed in the $c$-polarized 
spectra of the Bi2212 single crystal (T = 150 K, see Figure 3 and Table 6). 
Mixing due to out-of-phase motion of ions of equal symmetry in inequivalent
positions often does not allow the determination of a unique character for a
given set of equal symmetry modes. The total number of the modes in each
symmetry, of course, is unaffected by mixing. The character of the assigned
modes is determined according to the dominant contribution from the atoms
participating in the vibration and shown atop. The respective calculated
components of the eigenvector amplitudes are given in Table 9. 
The lowest frequency mode
calculated at 93 cm$^{-1}$ (experimentally observed at 97 cm$^{-1}$, see
Figure 3 and Ref. \cite{Zelezny1}) is related to motion of the heavy Bi
atoms, which is strongly mixed with out-of-phase motion of Sr, Cu1 and Ca,
with a small oxygen contribution from different crystallographic planes. The
next mode of A$_{2u}$ symmetry calculated at 195 cm$^{-1}$ (168 cm$^{-1}$) 
is predominantly due to motion of the Cu1 against Sr, with a
small contribution from the Ca ions. 
The A$_{2u}$ mode at 304 cm$^{-1}$ (210 cm$^{-1}$) is due
to the vibrations of Ca and O1 ions against Sr and O3. 
This mixed low-frequency mode 
contains noticeable contribution from the copper-plane O1 ions and 
bismuth-plane O3 ions to the $c$ axis oscillations. 
The three high-energy vibrations are predominantly oxygen-related. The mode
calculated at 343 cm$^{-1}$ (304 cm$^{-1}$) is basically due to 
O3 and O1 vibrations against Sr and Ca atoms, with a predominant 
contribution from the bismuth-plane oxygen O3. 
The Cu-O bond-bending character mode at 403 cm$^{-1}$ (358 cm$^{-1}$) 
involves significant contribution from the copper-plane O1
ions, vibrating against the Ca and O3 atoms. 
The highest frequency mode at 643 cm$^{-1}$ (583 cm$^{-1}$) is due to O2 and O3
out-of-phase oscillations, and it has apparently apical character due to
predominant contribution from the strontium-plane O2 oscillations along the $c$ axis. 
Our assignment for the $c$-polarized phonon modes is confirmed by 
the oxygen isotope effect in the Bi2212 single crystal, 
investigated in this work (see Figure 5). A noticeable red shift 
upon $^{18}$O substitution is observed for the four $c$ axis 
phonons at 210, 304, 358, and 583 cm$^{-1}$; the estimated values of the shifts
are 4.1, 4.4, 5.3, and 5.5\%, respectively. This result is consistent with 
our assignment that suggests that the three high-frequency modes 
have predominantly oxygen character. In addition, according to our assignment, 
the low-frequency mode at 210 cm$^{-1}$ also contains appreciable contribution 
from the copper-plane O1 and bismuth-plane O3 atoms, thus consistent with
the noticeable red shift with the $^{16}$O for $^{18}$O substitution. 
The present theoretical shell model calculations thus reproduce the
frequencies of the c axis A$_{2u}$ eigenmodes quite well for the bilayer
Bi2212 single crystal. We would like to notice that although 
the eigenfrequencies are somewhat different the character 
of A$_{2u}$ vibrations resulting from the present shell model calculations 
for the Bi2212 system is quite consistent with the earlier calculations, 
made by Prade {\it et al.} \cite{Prade}. 
 
In Figure 8 we show the eigenvector patterns corresponding to the eight
eigenmodes of A$_{2u}$ symmetry and two silent modes of B$_{2u}$ symmetry
for the Bi2223, resulting from our lattice-dynamical calculations
with the same set of the shell model parameters listed in Table 7. 
The calculated components of the eigenvector amplitudes are summarized in Table 9. 
We make an assignment of the TO phonon modes observed in the $c$ axis 
ellipsometry spectra of the Bi2223 single crystal (T = 150 K, see Figure 3 and Table 6) 
based on a comparison with the present calculations. As can be seen from 
Figures 7 and 8, close similarity with the Bi2212 is observed for their
eigenmode characters and eigenfrequency values due to the common generic
features in the Bi2212 and Bi2223 structures.
The only exception is that two new additional modes of A$_{2u}$ symmetry 
are present for the trilayer Bi2223 system. 
Our calculations give one additional mode for Bi2223 at 145 cm$^{-1}$, 
its character is defined by the main contribution from the Cu2 
(from the inner CuO$_2$ plane) and Sr out-of-phase vibrations, 
and assigned it with the new mode clearly discernible in 
 the $\epsilon_1(\omega)$ spectrum at 127 cm$^{-1}$ (see Figure 3(b)), 
which is also in agreement with the results of IR
measurements done very recently on an oriented ceramics of Bi2223 \cite{Petit}. 
We calculated four A$_{2u}$ "light", predominantly oxygen-related,
modes at the following frequencies 335, 378, 516, and 666 cm$^{-1}$. Like in
the Bi2212, the highest-frequency mode at 666 cm$^{-1}$ (581 cm$^{-1}$) is associated with
the out-of-phase O2 and O3 vibrations, the mode calculated at 335 cm$^{-1}$ (304 cm$^{-1}$)
is predominantly due to the Bi-plane O3 oscillations, and the mode 
calculated at 378 cm$^{-1}$ (359 cm$^{-1}$) is the Cu-O bond-bending character mode, 
corresponding to the out-of-phase motion of the Ca atoms and oxygens O1 in
the outer CuO$_2$ planes. We calculated one additional oxygen-related mode in
the Bi2223 at 516 cm$^{-1}$, and its character is defined by the out-of-phase 
vibrations of the Ca atoms and oxygens O4 in the inner CuO$_2$ plane. 
We assign it with the new mode appearing in $c$ axis the phonon spectrum of 
the Bi2223 at 402 cm$^{-1}$, as it is clearly observable in Figure 3, 
and also in agreement with the results of \cite{Petit}.

As we discussed in Section 4 some of the $c$-polarized IR phonon modes 
in the bilayer Bi2212 and trilayer Bi2223 compounds exhibit 
a strongly anomalous temperature dependence at T$_c$, accompanied with 
the development of a sizable absorption peak below T$_c$ in the FIR range, 
as illustrated in Figure 6 \cite{Munzar1,Zelezny1,Munzar2,Boris}. 
The two most pronounced phonon bands at 304 and 582 cm$^{-1}$ appear 
at the same frequencies in the $c$ axis ellipsometry spectra of the Bi2212 and Bi2223. 
According to our assignment they are associated with the predominant contribution from the BiO layer O3 oxygens and 
the apical O2 oxygens, respectively. 
It is evident from Figure 6 that the formation of the t-JPR in Bi2212 and Bi2223 
is accompanied with the strongly anomalous temperature dependence of 
the phonon modes at 360, 402 (specific for trilayer Bi2223) and 582 cm$^{-1}$ 
denoted by $A, B, C$. An explanation of the anomalous changes 
of the phonon modes at 360 and 402 cm$^{-1}$ upon entering the superconducting
state has been recently proposed based on the so-called Josephson superlattice model,
assuming that these bands belong to the vibration of the planar oxygens \cite{Munzar2,Boris}. 
This assumption is supported by our assignment which suggests that these modes are bond-bending character modes 
which involve a significant contribution from the outer copper-plane O1 
and inner copper-plane O4 ions, respectively.
\begin{table}[tbp]
\caption{Calculated components of eigenvectors of A$_{2u}$ modes in Bi2201
(Bi$_2$Sr$_2$CuO$_6$), Bi2212 (Bi$_2$Sr$_2$CaCu$_2$O$_8$)  and Bi2223 (Bi$_2$Sr$_2$Ca$_2$Cu$_3$O$_{10}$) 
crystals in tetragonal \textit{I4/mmm} (D$^{17}_{4h}$) space group.}
\begin{ruledtabular}
\begin{tabular}{rrrrcrrrrc}
Bi2212    & Bi($z$)& Sr($z$)& Cu1($z$)& -- & Ca($z$) & O1($z$) & O2($z$)& O3($z$)& --\\    
\colrule
\ 93 $cm^{-1}$ & -0.46 & 0.27  & 0.33 & --    & 0.29    & 0.16  & -0.07& -0.11  & -- \\
195 $cm^{-1}$  & -0.04 &  0.43 & -0.54& --    &  0.16   & 0.05  & -0.01&  0.00  & -- \\
304 $cm^{-1}$  & -0.09 &  0.33 &  0.10& --    & -0.76   &-0.15  &  0.08&  0.18  & -- \\
 343 $cm^{-1}$  & -0.11 & -0.18 & -0.09& --    & -0.15   & 0.31  &  0.13&  0.46  & -- \\
403 $cm^{-1}$  &  0.09 & -0.01 & -0.04& --    & -0.50   & 0.29  & -0.15& -0.41  & -- \\
643 $cm^{-1}$  &  0.07 &  0.04 &  0.03& --    &  0.01   & -0.01 & -0.66&  0.24  & -- \\
\colrule
Bi2223      & Bi($z$)& Sr($z$)& Cu1($z$)& Cu2($z$)& Ca($z$) & O1($z$) & O2($z$)& O3($z$)& O4($z$)\\    
\colrule
\ 88 $cm^{-1}$ &-0.45  & 0.11 & 0.14 &  0.57 &  0.21  & 0.08   & -0.07 & -0.11  & 0.17  \\
145 $cm^{-1}$  &-0.23  & 0.33 & 0.22 & -0.68 &  0.09  & 0.10   &  0.00 & -0.01  & -0.09 \\
211 $cm^{-1}$  & 0.01  & 0.27 & -0.59& -0.06 &  0.23  & 0.06   & -0.03 &  0.00  & 0.09  \\
252 $cm^{-1}$  & -0.05 & 0.42 & -0.05&  0.29 & -0.49  & -0.07  &  0.02 &  0.03  & -0.16 \\
335 $cm^{-1}$  & -0.09 & -0.15& -0.07& -0.01 & -0.21  &  0.29  &  0.10 &  0.46  & 0.16  \\
378 $cm^{-1}$  &  0.10 & -0.09& -0.05&  0.09 & -0.13  &  0.34  & -0.14 & -0.41  & 0.04 \\
516 $cm^{-1}$  & 0.05  & 0.05 & 0.03 & -0.21 & -0.23  & -0.05  & -0.08 & -0.22  & 0.60  \\
666 $cm^{-1}$  & 0.08  & 0.03 & 0.05 &  0.00 & 0.01   & -0.01  & -0.66 & 0.22   & 0.02 \\
\colrule
Bi2201      & Bi($z$)& Sr($z$)& Cu1($z$)& -- & -- & O1($z$) & O2($z$)& O3($z$)& --\\    
\colrule
\ 77 $cm^{-1}$&-0.31 &  0.24  &  0.77 & -- & --  &  0.10  & -0.11 & -0.17  & -- \\
175 $cm^{-1}$ &-0.21 &  0.51  & -0.56 & -- & --  &  0.17  &  0.02 & -0.06  & -- \\
293 $cm^{-1}$ &-0.24 & -0.09  &  0.08 & -- & --  &  0.30  &  0.15 &  0.56  & -- \\
353 $cm^{-1}$ & 0.10 & -0.23  & -0.08 & -- & --  &  0.59  & -0.10 & -0.27  & -- \\
596 $cm^{-1}$ & 0.09 &  0.06  & -0.05 & -- & --  &  0.00  & -0.66 &  0.23  & -- \\
\end{tabular}
\end{ruledtabular} 
\end{table}

In Figure 9 we present the eigenvector pattern corresponding to the five
eigenmodes of A$_{2u}$ symmetry for Bi2201, the calculated components 
of the eigenvector amplitudes are listed in Table 9. 
We have calculated two A$_{2u}$ symmetry modes at low frequencies. 
The calculated lowest frequency mode is related to the motion of the heavy Bi atoms, 
which is strongly mixed with out-of-phase motion of Sr and Cu1. 
The next mode of A$_{2u}$ symmetry is predominantly due to the motion 
of Cu1 against Sr. The three phonon features are clearly observed in 
the $c$ axis ellipsometry spectra of the Bi2201 at low frequencies at around 
109, 165, and 209 cm$^{-1}$ (see Figure 4 and Table 6); in Section 4 we have already 
discussed the possible reasons for the appearance of one "extra" phonon feature.
We have made an assignment of the modes at low frequencies by analogy with 
the vibration characters at the corresponding frequencies in the B2212 and Bi2223 crystals. 
The three high-energy vibrations are predominantly 
oxygen-related. The highest frequency mode at 596 cm$^{-1}$ (586   cm$^{-1}$) 
has apparently apical character due to the predominant contribution 
from the strontium-plane O2 oscillations along the $c$ axis. 
The modes calculated at 
293 cm$^{-1}$ and 353 cm$^{-1}$ are basically due to bismuth-plane O3 
and copper-plane O1 mixed in-phase and out-of-phase vibrations, respectively, 
and they correlate well with the experimental frequencies at 301 and 385 cm$^{-1}$. 
In the Bi2201 the CuO$_2$ planar oxygen band located at 385 cm$^{-1}$ has
much larger intensity than in the Bi2212. This is very probable due to
the fact that in the Bi2212 the CuO$_2$ planar oxygen mode is, already
in the normal state, strongly renormalized by its coupling to interlayer
electronic excitations.

\subsection{Calculated $\Gamma$-point A$_{1g}$ symmetry modes in Bi2201,
Bi2212 and Bi2223}

In this Section we just briefly discuss the results of our shell model
calculations for the Raman-active modes in the three Bi-based compounds and
suggest the assignment for the phonon modes observed in the experimental
spectra. It has been generally found that a larger number of phonon modes 
occur in the Raman spectra of the Bi-based compounds than would be expected 
from the simple tetragonal $I4/mmm$ symmetry consideration (given in Tables 1 and 2). 
As a result different groups have reported rather different and then contradictory assignments.
We have already referred to these reasons in Section 2. There appears to be a
consensus on the origin of the strong B$_{1g}$ phonon in Raman spectra of
Bi2212. From the B$_{1g}$-symmetry selection rules the peak at about 285 cm$^{-1}$, 
which occurs only in the $z(xy)\bar z$ spectrum can be assigned to the
O(1) B$_{1g}$ phonon characterized by out-of-phase motion of the oxygen
atoms in the CuO$_2$ plane. This is not the case, however, for two of the
strongest A$_{1g}$ modes that appear in the Raman spectrum at higher
frequencies around 460 and 620 cm$^{-1}$. The origin of the high-frequency
vibrations in the Raman spectrum of Bi2212 remains controversial despite
many investigations by different groups. Most authors have assigned the 460
cm$^{-1}$ mode to the vibrations of the apical oxygen atoms located in the
SrO planes and the 620 cm$^{-1}$ phonon to vibrations of the oxygen atoms
located in the BiO planes \cite{Cardona,Liu,Boekholt}. However, recently it
was argued that the assignment should be reversed \cite{Kakihana,Pantoja}.
It was shown that these two bands exhibit different polarization
dependencies for the incident and scattered vectors, indicating that the
degree of mode mixing is small, and the 620 cm$^{-1}$ band is strongly $c$
-polarized like apex oxygen vibrations characteristic for most high-$T_c$
superconductors \cite{Kakihana}. Moreover, the 460 cm$^{-1}$ band has been
found to soften at a significantly faster rate than the 620 cm$^{-1}$ band
with $^{18}$O oxygen isotope substitution, in agreement with the more liable
oxygen atoms from the BiO layer \cite{Pantoja}. Therefore, in fitting our
shell model parameters for the Bi2212 crystal we have implied an assignment
for the mode observed at around 285 cm$^{-1}$ as the B$_{1g}$ mode, whereas
for the modes observed at around 460 and 620 cm$^{-1}$ as the A$_{1g}$ modes
associated with the O3 (from the BiO plane) and O2 (from the SrO plane)
predominant contributions, respectively.

In Figure 10 we show our suggested eigenvector pattern corresponding to the
six eigenmodes of A$_{1g}$ symmetry and one B$_{1g}$ symmetry for 
the Bi2212 crystal. We list values of the eigenfrequencies and below 
we make an assignment in agreement with the experimental Raman study of 
Bi$_2$Sr$_2$Ca$_{1-x}$Y$_x$Cu$_2$O$_{8+\delta}$ ($x$ = 0$\ \div$\ 1) 
single crystals \cite{Kakihana}. A number of phonons observable below 400 cm$^{-1}$ 
in the Raman spectra in the A$_{1g}$ (or A$_g$) scattering configurations 
are considered by Kakihana {\it et al.} as "disorder-induced" modes \cite{Kakihana}. 
In addition, as a weak satellite feature at $\sim$ 657 cm$^{-1}$ near 
the strong phonon band at 620 cm$^{-1}$ disappears in modulation-free 
Bi$_{2-x}$Pb$_x$Sr$_2$Ca$_{1-x}$Cu$_2$O$_{8+\delta}$ crystals it is possibly 
induced by the incommensurate superstructural modulation \cite{Ben}. 
Another satellite feature at $\sim$ 458 cm$^{-1}$ near the pronounced 
phonon band at 460 cm$^{-1}$ is possibly associated with extra oxygen atoms 
in the BiO layers, and also referred to "disorder-induced" scattering. 

In Figure 11 we show the eigenvector pattern corresponding to the seven
eigenmodes of A$_{1g}$ symmetry and one B$_{1g}$ symmetry for the Bi2223 crystal, 
calculated using the same basic set of the shell model parameters. 
From Figures 10 and 11 we can see that due to the common generic features 
in the Bi2212 and Bi2223 structures the vibration characters are similar and 
the eigenfrequency values are close for the corresponding eigenmodes. 
This is consistent with the recently published 
results on Raman measurements showing that frequencies of the major phonon 
bands are almost identical in Bi2212 and Bi2223 single crystals \cite{Limonov}. 
The exception is an additional mode of the A$_{1g}$ symmetry present in the
trilayer Bi2223 crystal. Comparing the bilayer and trilayer systems (Figures
10 and 11) we can conclude that instead of one A$_{1g}$ mode associated with
the O1 vibrations in the CuO$_2$ plane in the Bi2212 structure, calculated
at 370 cm$^{-1}$, in the Bi2223 structure we have two A$_{1g}$ modes
associated with the Ca and O1 (from the outer CuO$_2$ planes) in-phase and
out-of-phase vibrations, calculated at 316 and 391 cm$^{-1}$, respectively.

In Figure 12 we show the eigenvector pattern corresponding to the four
eigenmodes of A$_{1g}$ symmetry for the $I4/mmm$ Bi2201, 
resulting from our shell model calculations. As usual many more phonon modes 
are observed in the experimental Raman spectra implying that the simple 
tetragonal approximation is no more applicable for the interpretation 
of the Raman phonon spectra of the Bi2201. In Tables 1 and 3 we show that due 
to the orthorhombic distortions the O1 atoms in the CuO$_2$ plane 
become "Raman-active", and, as a result, a number of new Raman-active modes of 
A$_g$, B$_{1g}$, 2B$_{2g}$, and 2B$_{3g}$ character will appear in the $Amaa$ Bi2201. 
We suggest that a more appropriate approximation for the interpretation of 
the $\Gamma$-point Raman-active phonons in this compound will be 
the orthorhombic $Amaa$ space group. However, the assignment 
for the two high-frequency oxygen modes (as we show in Figure 12) is the same 
as for the Bi2212 and Bi2223 crystals, and the calculated frequencies 
agree quite well for these three compounds.

Resonant Raman scattering below T$_c$ has been discovered in Bi- and Hg-based 
high T$_c$ compounds with three or four CuO$_2$ layers \cite{Hadjiev,Limonov,Munzar3}.  
Strong superconductivity-induced enhancement of the phonons at 260 and 390 cm$^{-1}$, 
together with the development of a broad band peaking at around 580 cm$^{-1}$, 
has been observed in the recent resonant Raman scattering study on the
Bi2223 single crystal with T$_c$ = 109 K \cite{Limonov}. 
Our measurements of resonant Raman scattering (E$_{exc}$ = 2.18 eV) done on our Bi2223 single
crystal with lower T$_c$ = 107 K reproduce the results of Limonov {\it et al.} 
In Figure 13(a) we show the Raman spectra at two different temperatures, 
i.e. somewhat above and well below T$_c$. Figure 13(b) shows the difference between these Raman
spectra. Our data clearly demonstrate the superconductivity-induced enhancement 
of the phonon bands at 260 and 390 cm$^{-1}$ and a broad  band peaking around 550 cm$^{-1}$. 
There are two alternative explanations of these superconductivity-induced anomalies 
in multilayer high-T$_c$ compounds; one considers the broad excitation band as a "pair-breaking"
peak at 2$\Delta$ \cite{Limonov}, and the another refer it to a Raman-active $c$ axis plasmon \cite{Munzar3}.
A more detailed study of the doping effect on the resonance peak position
is required to clarify its origin. Nevertheless, dicussing the phonon anomalies
observed in trilayer Bi2223 we would like to notice that both approaches
consider the superconductivity-induced anomalies as due to strong coupling
of the intratrilayer electronic excitation with phonons which involve calcium 
and planar oxygen ions. Experimental resonant Raman scattering studies bring evidence in favour of
our assignment suggested for the trilayer Bi2223 crystal (Figure 11). 
The superconductivity-enhanced phonon bands at 260 and 390 cm$^{-1}$
in the Bi2223 can be ascribed to the A$_{1g}$ in-phase and out-of-phase vibrations
of the Ca atoms with oxygens in the outer CuO$_2$ planes, 
in agreement with our assignment for the A$_{1g}$ modes calculated at 316
and 391 cm$^{-1}$. 

\section{Conclusions}

In summary, we present experimental data on single crystals 
as obtained by spectral ellipsometry and make the assignment of the 
IR $c$ axis polarized phonon modes in single, bi-, and trilayer 
bismuth compounds (Bi2201, Bi2212, and Bi2223) by comparing them 
with the results of the shell model calculations. 
The $c$ axis IR phonon spectra look very similar for the bilayer and trilayer 
Bi-based crystals, supporting experimental evidence for the generic structural features 
of these compounds. We apply a tetragonal $I4/mmm$ approximation 
in the interpretation of the $c$ axis IR spectra of the Bi-based compounds, 
which are known to have pseudotetragonal structure, 
and conclude that this approximation provides a reasonable description of the
main features in the phonon spectra.  
In the framework of this approximation, we have identified "disorder-induced" 
modes due to the orthorhombic distortions associated with 
the incommensurate superstructural modulations and extra-stoichiometric oxygen. 
Comparing with the bilayer Bi-system, we identify two additional $c$ axis phonon 
modes in the ellipsometric spectra of the trilayer system, 
one at 127 cm$^{-1}$, among the "heavy" modes, and the another at 402 cm$^{-1}$, 
among the "light" modes, which according to our assignment 
involve significant contribution of the copper and oxygen atoms 
in the inner CuO$_2$ plane, respectively. 
The recent experimental observation of a Josephson-plasma-resonance 
and phonon anomalies in Bi2223 \cite{Boris} provide, indeed, a strong support 
for the Josephson-superlattice model, which consistently explains, 
the phonon anomalies and the formation of a broad absorption band below $T_c$ 
using the components of the eigenvector amplitudes for the relevant oxygen modes 
at 358 and 402 cm$^{-1}$. The two modes involving O3 (BiO plane) and O1 (CuO$_2$ plane) 
atoms exhibit substantial frequency changes in the single layer system, and appear at
294 and 389 cm$^{-1}$, respectively. For the single layer system the mode at 389
cm$^{-1}$ increases dramatically in intensity. This evident difference 
between single and multilayer compounds in the behavior of this particular phonon mode  
reflects the fact that in multilayer compounds the oxygen bond-bending phonon mode 
is strongly-coupled with the electronic excitation due to the charge 
density oscillations between the closely spaced CuO$_2$ planes. 
In addition, we also show the eigenvector patterns and calculated frequencies
for the Raman-active phonon modes of A$_{1g}$ symmetry in these three 
compounds according to the results of the present shell model calculations. 
The supeconductivity-induced resonant Raman scattering for the two phonon
modes at 260 and 390 cm$^{-1}$ very recently observed in the trilayer Bi2223 
single crystal \cite{Limonov} is well consistent with the suggested assignment.

\section{Acknowledgments}
The authors thank J. Gale for making available General Utility Lattice 
Program (GULP) used in the present calculations. 
We are grateful to R. Evarestov and E. Kotomin for fruitful discussions. 
T. Holden acknowledges support by AvH Foundation. 
We acknowledge L. Carr at NSLS and Y.-L. Mathis at ANKA for their support
during the ellipsometry measurements.

\newpage

\begin{figure}[tbp]
\includegraphics*[width=130mm]{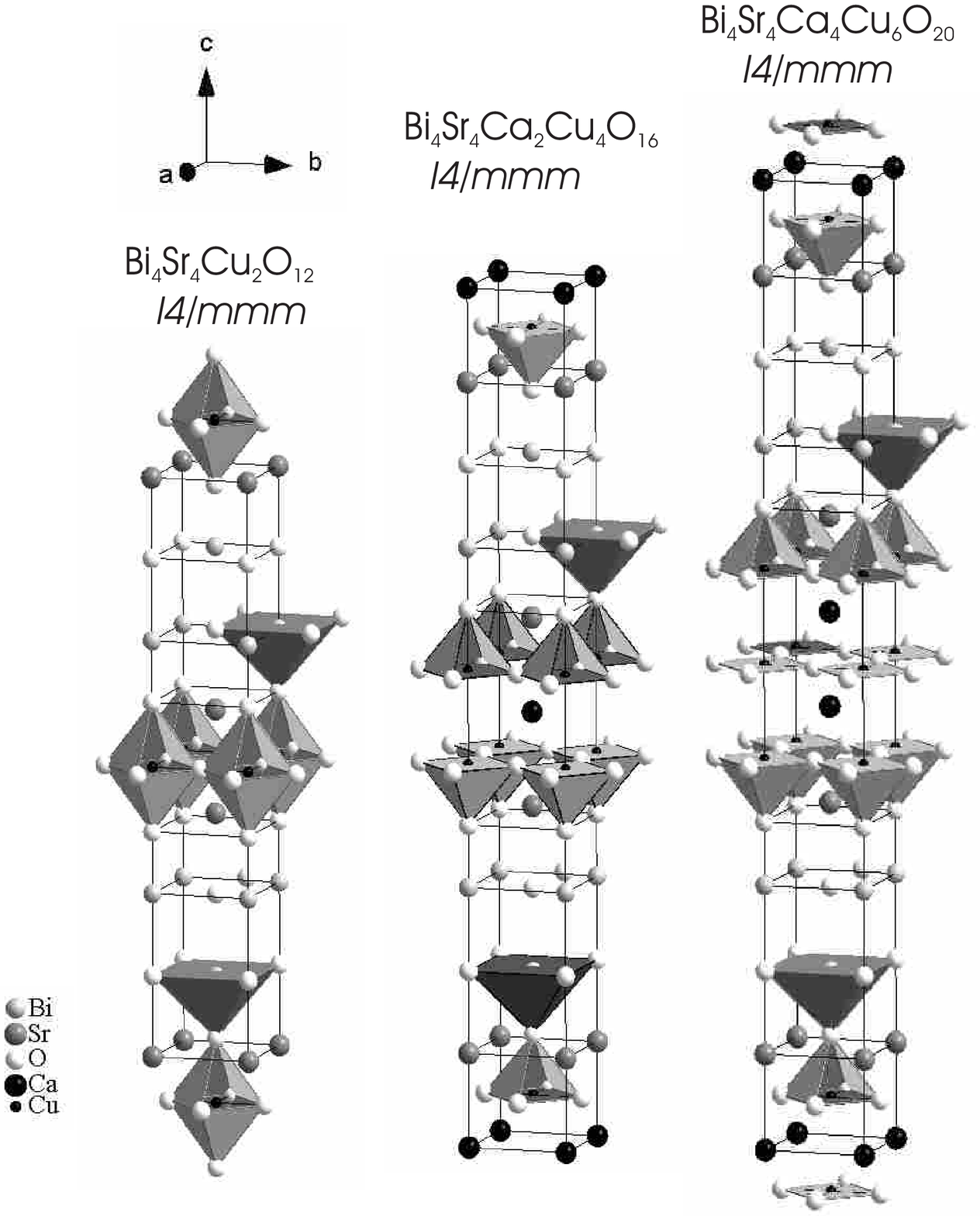}
\caption{Crystallographic cells of Bi$_2$Sr$_2$Ca$_{n-1}$Cu$_n$O$_{2n+4}$ with n = 1, 2 and 3.} 
\label{Fig1}
\end{figure}

\begin{figure}[tbp]
\includegraphics*[width=100mm]{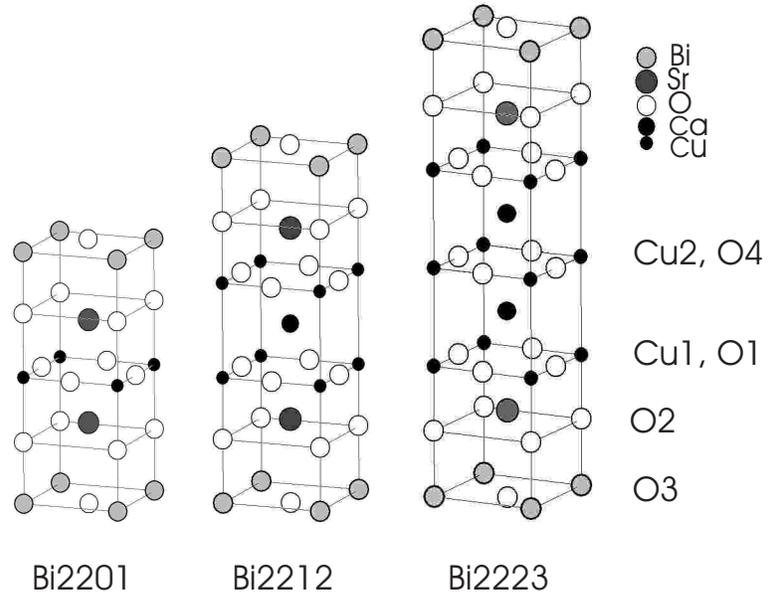}
\caption{Notations for copper and oxygen atoms in primitive unit cells 
of single, bi-, and trilayer Bi-based cuprates.}
\label{Fig2}
\end{figure}

\begin{figure}[tbp]
\includegraphics*[width=140mm]{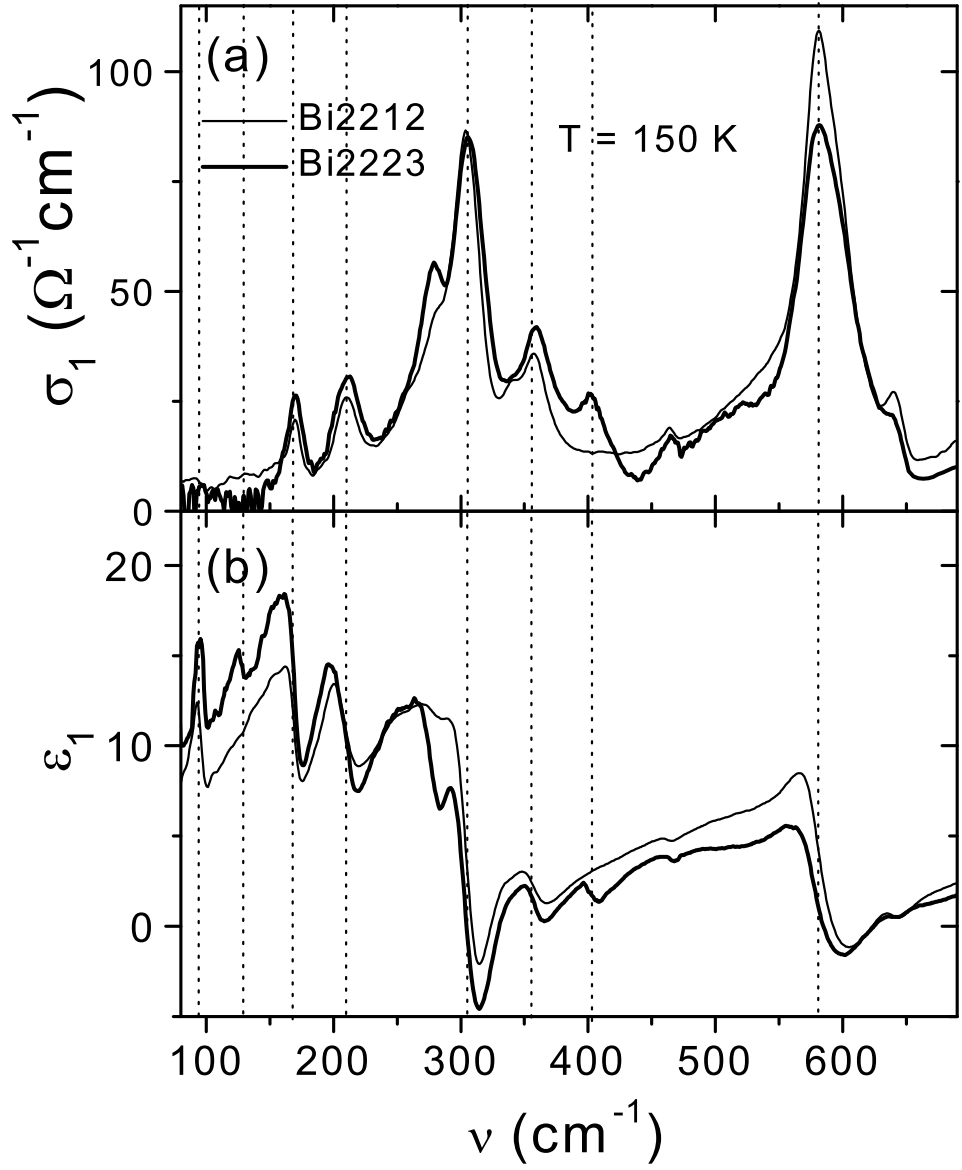}
\caption{Real parts of the $c$ axis (a) optical conductivity $\sigma_1(\omega)$ 
and (b) dielectric function $\epsilon_1(\omega)$ of Bi2212 and Bi2223 single 
crystals with T$_c$ = 91 K and T$_c$ = 107 K at T = 150 K.}
\label{Fig3}
\end{figure}

\begin{figure}[tbp]
\includegraphics*[width=140mm]{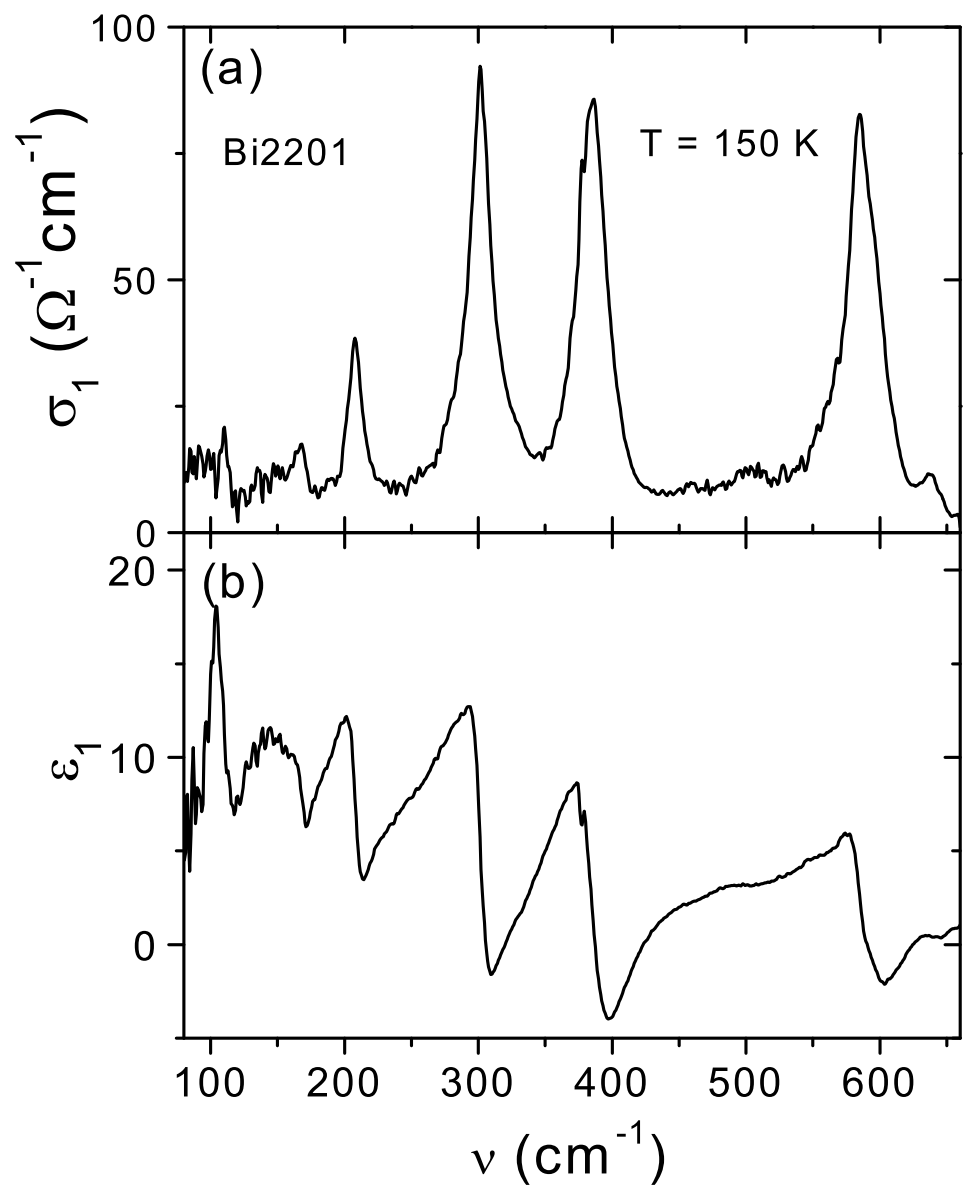}
\caption{$C$ axis (a) optical conductivity $\sigma_1(\omega)$ 
and (b) dielectric function $\epsilon_1(\omega)$ of Bi2201 single crystal at T = 150 K.}
\label{Fig4}
\end{figure}

\begin{figure}[tbp]
\includegraphics*[width=140mm]{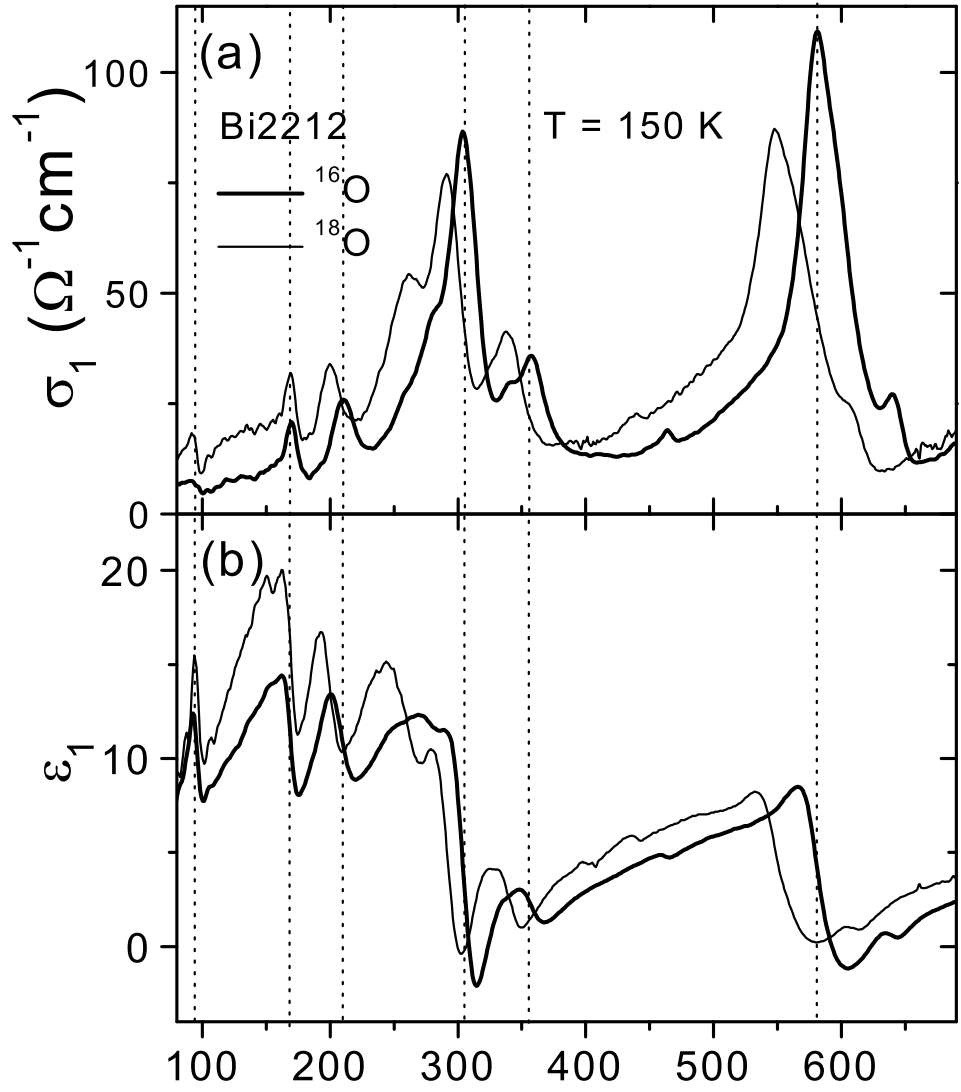}
\caption{Oxygen isotope effect in the $c$-polarized response of (a) $\sigma_1(\omega)$ 
and (b) $\epsilon_1(\omega)$ of Bi2212 with T$_c$ = 91 K, measured at 150 K.}
\label{Fig5}
\end{figure}

\begin{figure}
\includegraphics*[width=130mm]{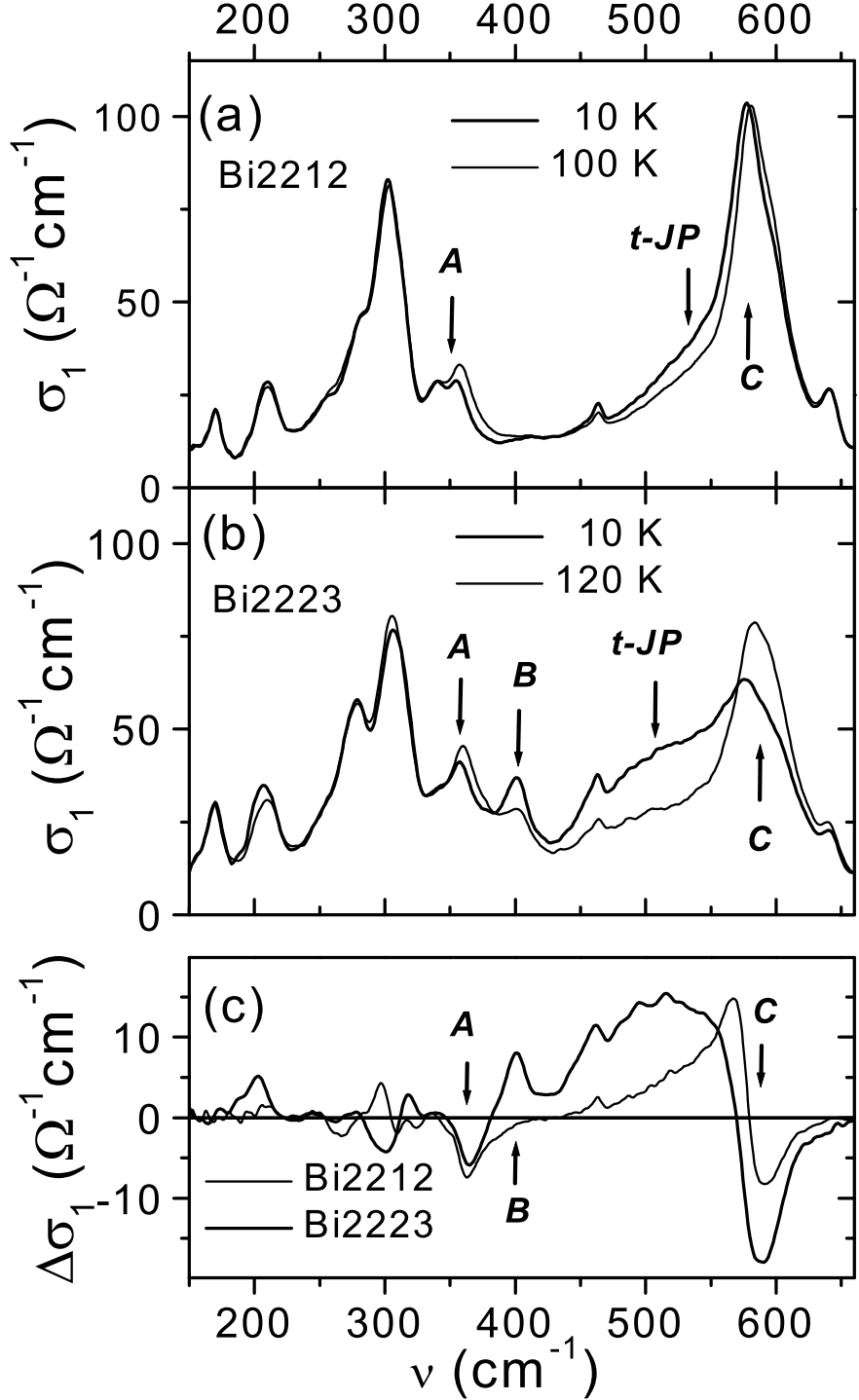}
\caption{Real part $\sigma_{1}(\omega)$ of the FIR $c$ axis conductivity
(a) of Bi2212 with $T_c$ = 91 K for $T$ = 100 K and $T$ = 10 K \cite{Munzar2} and 
(b) of Bi2223 with $T_c$ = 102 K for $T$ = 120 K and $T$ = 10 K \cite{Boris}.
(c) Superconductivity-induced changes of the FIR conductivity,
$\Delta \sigma_1=\sigma_1 (T<<T_c,\omega) - \sigma_1(T\gtrsim
T_c,\omega)$, for Bi2212 and Bi2223 derived from the data shown in the top panels. 
The most pronounced phonon anomalies are denoted by $A$, $B$, and $C$.}
\label{Fig6}
\end{figure}

\begin{figure}[tbp]
\includegraphics*[width=160mm]{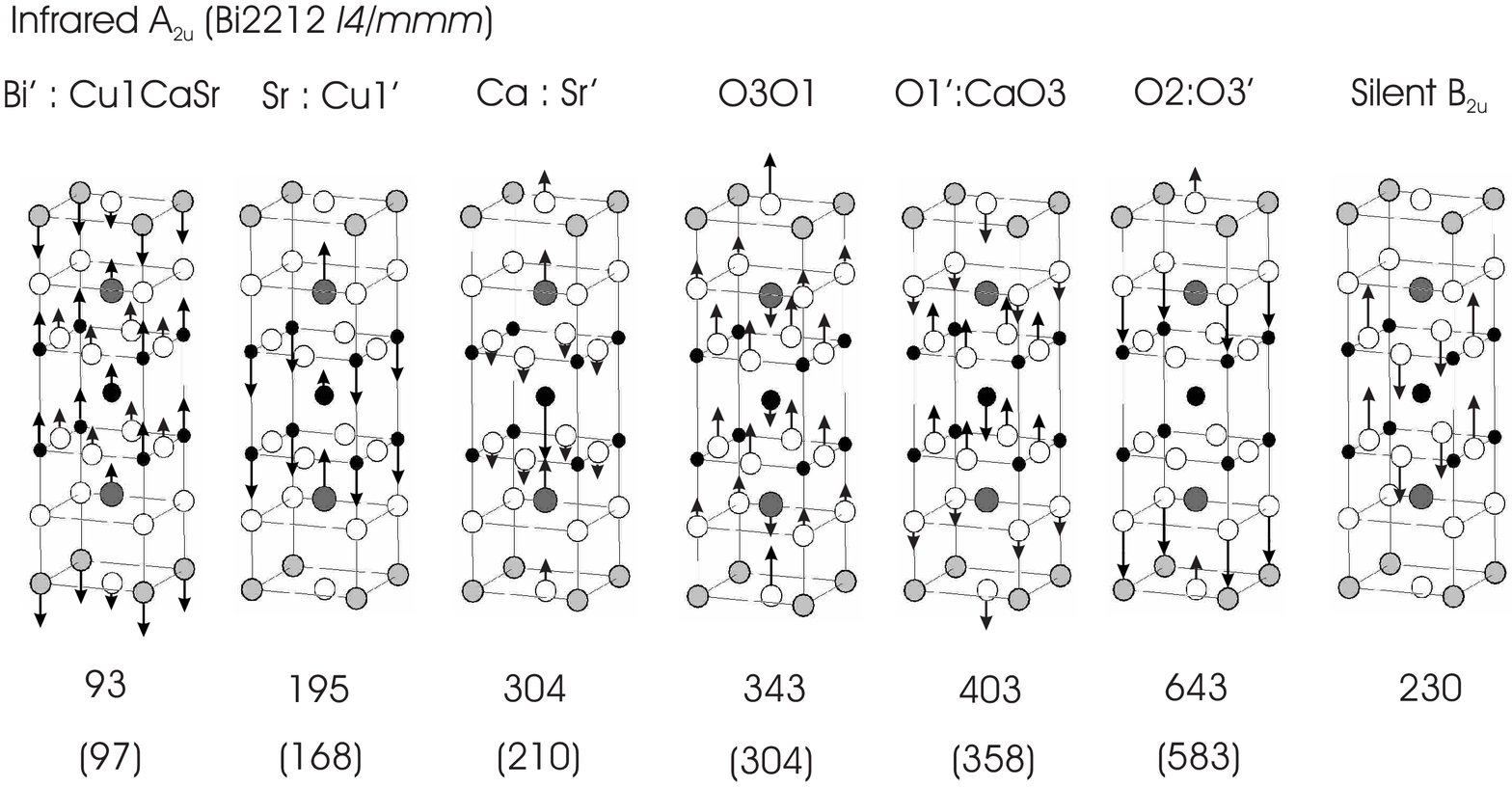}
\caption{Normal frequencies (in cm$^{-1}$) of six A$_{2u}$ symmetry modes and one "silent" B$_{2u}$ mode 
in Bi2212 ($I4/mmm$) and schematic representation of their eigenvectors; respective experimental 
$c$ axis TO phonon frequencies are given in brackets (see Figure 3 and Table 6).}  
\label{Fig7}
\end{figure}

\begin{figure}[tbp]
\includegraphics*[width=160mm]{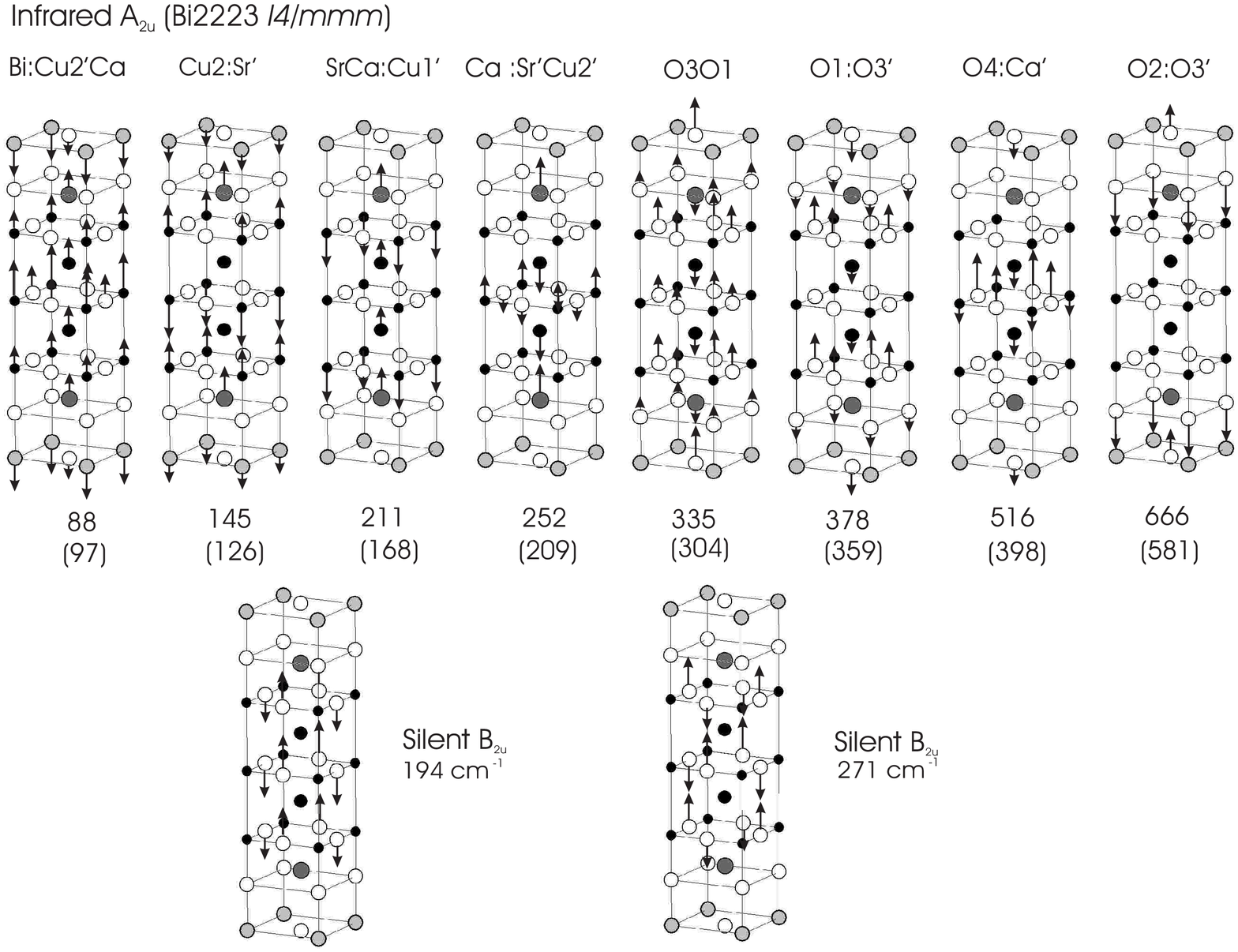}
\caption{Normal frequencies and eigenvectors of eight A$_{2u}$ symmetry modes and 
two "silent" B$_{2u}$ modes in Bi2223 ($I4/mmm$); respective experimental $c$ axis TO phonon 
frequencies are given in brackets (see Figure 3 and Table 6).}
\label{Fig8}
\end{figure}

\begin{figure}[tbp]
\includegraphics*[width=130mm]{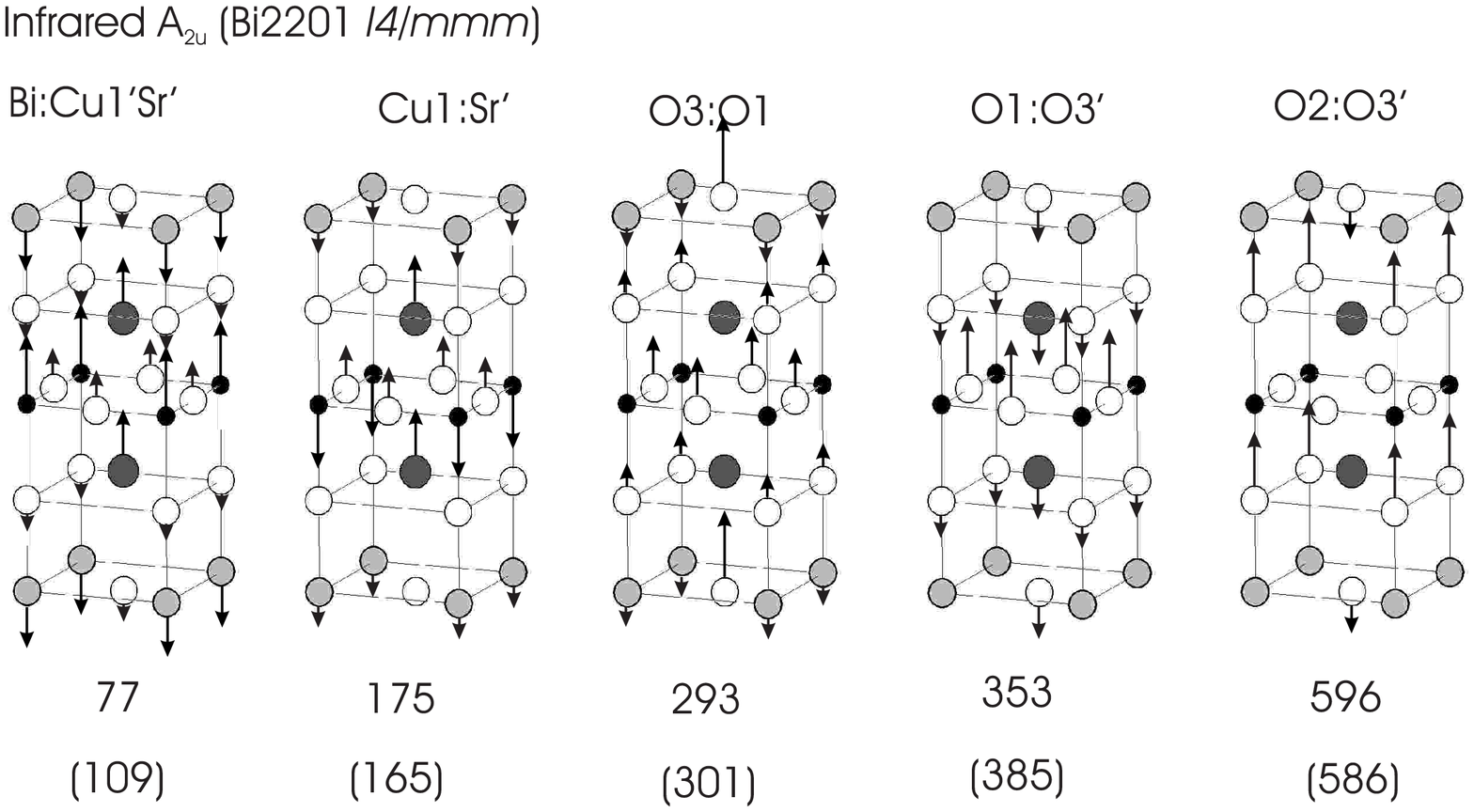}
\caption{Normal frequencies and eigenvectors of five A$_{2u}$ symmetry modes 
in Bi2201 ($I4/mmm$); experimental $c$ axis TO phonon frequencies are given 
in brackets (see Figure 4 and Table 6 ).} 
\label{Fig9}
\end{figure}
\newpage

\begin{figure}[tbp]
\includegraphics*[width=160mm]{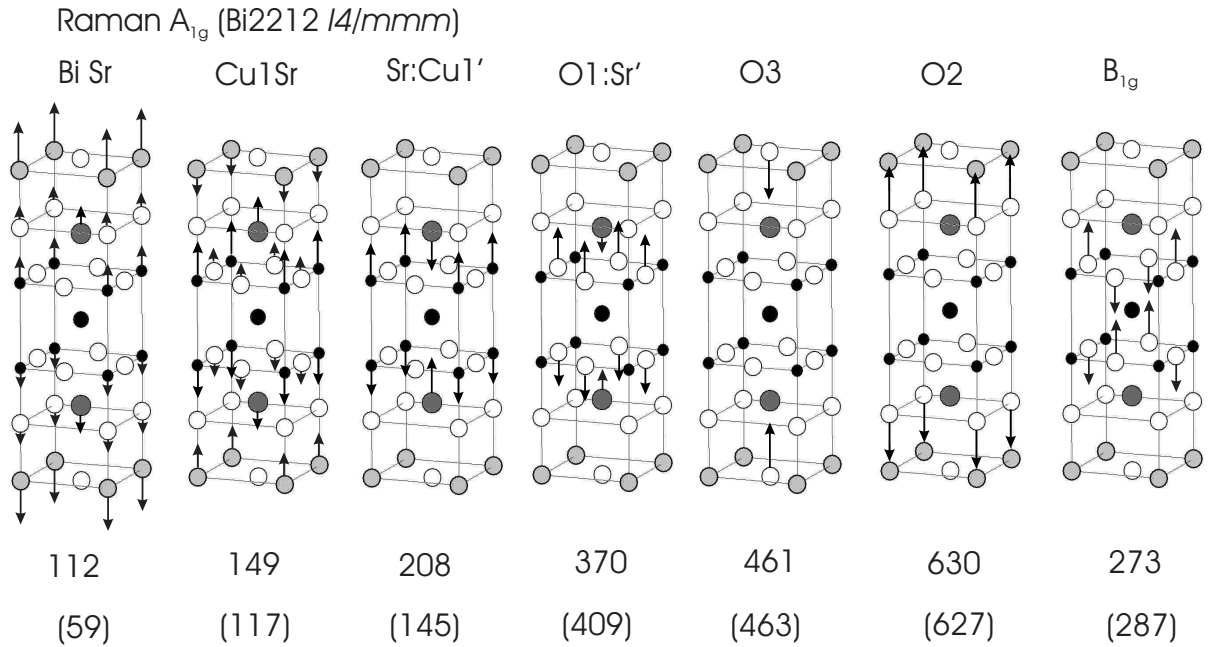}
\caption{Normal frequencies and eigenvectors of six A$_{1g}$ symmetry modes and
one B$_{1g}$ mode in Bi2212 ($I4/mmm$); experimental frequencies are given 
in brackets \cite{Kakihana}.}
\label{Fig10}
\end{figure}

\begin{figure}[tbp]
\includegraphics*[width=160mm]{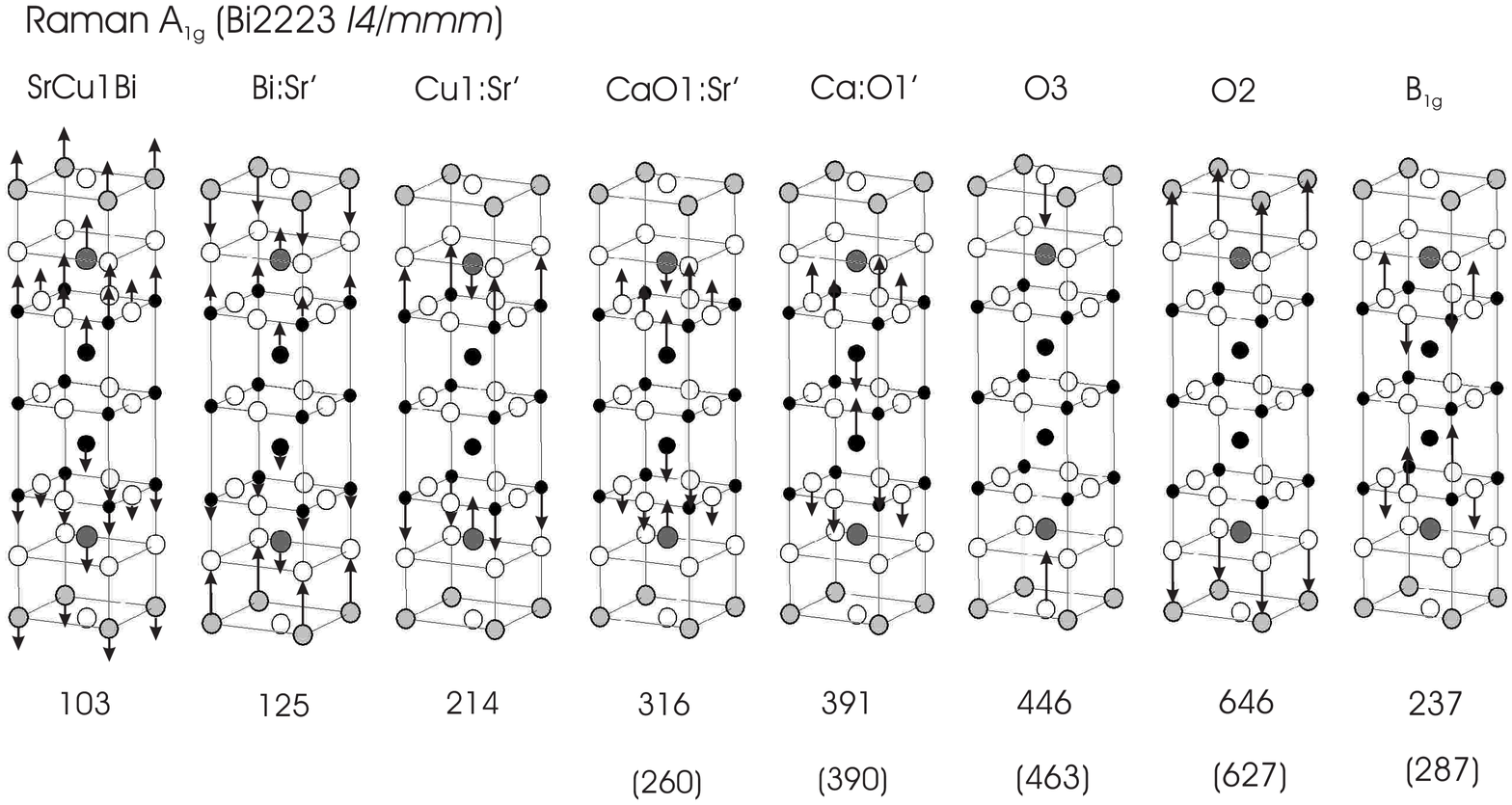}
\caption{Normal frequencies and eigenvectors of seven A$_{1g}$ symmetry modes and
one B$_{1g}$ mode in Bi2223 ($I4/mmm$); some of experimental frequencies are given 
in brackets (see Figure 13 and Ref. \cite{Limonov}).}
\label{Fig11}
\end{figure}

\begin{figure}[tbp]
\includegraphics*[width=120mm]{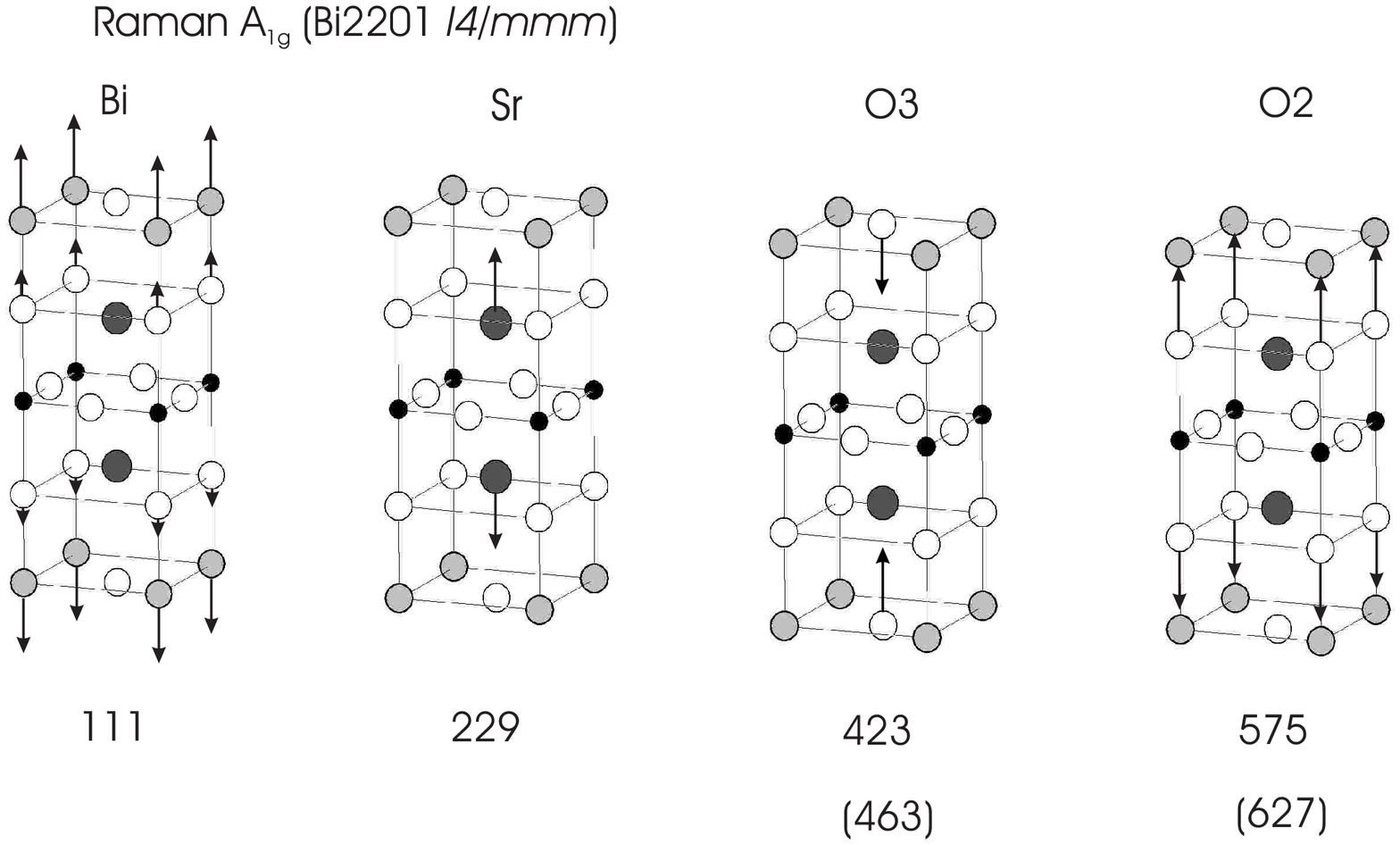}
\caption{Normal frequencies and eigenvectors of four A$_{1g}$ symmetry modes 
in Bi2201 ($I4/mmm$); some of the experimental frequencies are given in
brackets \cite{Kakihana}.}
\label{Fig12}
\end{figure}

\begin{figure}[tbp]
\includegraphics*[width=130mm]{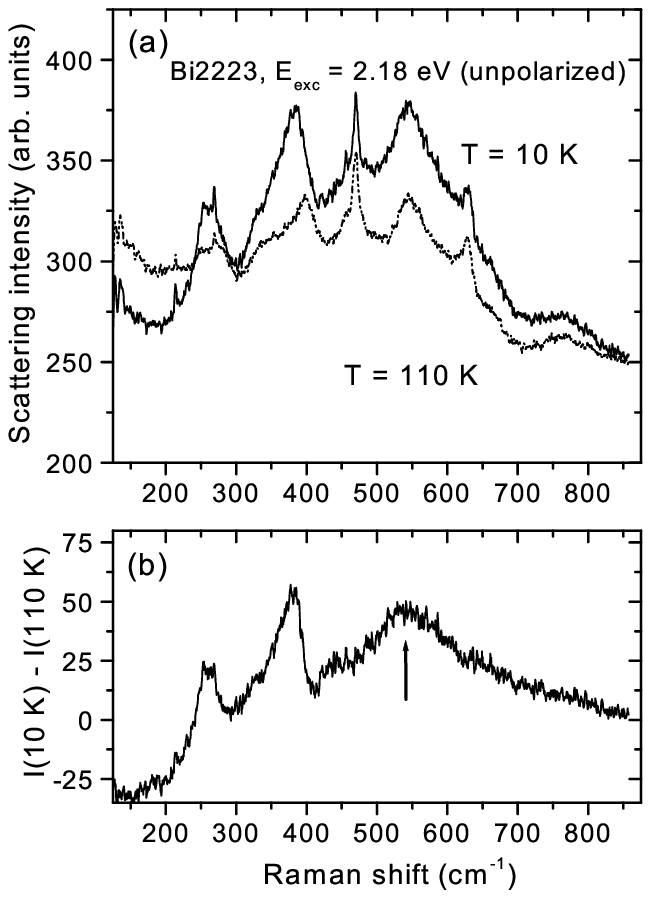}
\caption{Raman scattering ($E_{exc}$ = 2.18 eV) in Bi2223  
(a) measured slightly above T$_c$ = 107 K and well below T$_c$ and (b)
resonant enhancement of the phonon bands at 260 and 390 cm$^{-1}$ and 
superconductivity-induced broad band at 550 cm$^{-1}$.}
\label{Fig13}
\end{figure}


\begin{thebibliography}{99}
\bibitem{remark} Our earlier fit of the shell model parameters was based on
the experimental data only of the $c$-axis ir- phonons. Although these
earlier calculations resulted, in principal, in the same assignment and
character of the A$_{2u}$ modes as the present calculations, we arrived at
somewhat different eigenvalues of the A$_{2u}$ modes, in particular for the
Bi2223 presented in \cite{Boris}.
\end{thebibliography}
\end{document}